\documentclass[10pt]{article}
\usepackage{multicol}
\usepackage{graphicx}
\usepackage{amsmath}
\usepackage[a4paper]{geometry}

\begin{document}

\begin{center}
{\Large\bf Event patterns from negative pion spectra\\ in
proton-proton and nucleus-nucleus collisions at SPS}

\vskip0.5cm

Ya-Hui Chen$^{1}$, Fu-Hu Liu$^{1;}${\footnote{E-mail:
fuhuliu@163.com; fuhuliu@sxu.edu.cn}}, Edward K.
Sarkisyan-Grinbaum$^{2;3;}${\footnote{E-mail:
Edward.Sarkisyan-Grinbaum@cern.ch}}

{\small\it $^1$Institute of Theoretical Physics \& State Key
Laboratory of Quantum Optics and Quantum Optics Devices,

Shanxi University, Taiyuan, Shanxi 030006, China

$^2$Experimental Physics Department, CERN, 1211 Geneva 23,
Switzerland

$^3$Department of Physics, The University of Texas at Arlington,
Arlington, TX 76019, USA}
\end{center}

\vskip0.5cm

{\bf Abstract:} Rapidity-dependent transverse momentum spectra of
negatively charged pions measured at different rapidities in
proton-proton collisions at the Super Proton Synchrotron (SPS) at
various energies within its Beam Energy Scan (BES) program are
investigated by using one- and two-component standard
distributions where the chemical potential and spin property of
particles are implemented. The rapidity spectra are described by a
double-Gaussian distribution. At the stage of kinetic freeze-out,
the event patterns are structured by the scatter plots in the
three-dimensional subspaces of velocity, momentum and rapidity.
The results of the studies of the rapidity-independent transverse
mass spectra measured at mid-rapidity in proton-proton collisions
are compared with those based on the similar transverse mass
spectra measured in the most central beryllium-beryllium,
argon-scandium and lead-lead collisions from the SPS at its BES
energies.
\\

\\


\vskip1.0cm

\begin{multicols}{2}

{\section{Introduction}}

The spectra of transverse momentum and (pseudo)rapidity of
particles produced in high energy collisions are of high interest
as soon as they provide us with an important information of the
kinetic freeze-out state of the interacting system. The transverse
momentum spectrum, characterizing the system developments in the
transverse plane, is associated either with the effective
temperature or with the freeze-out temperature, where the first
contains contributions of thermal motion and flow effect, while
the latter is believed to get the contribution of the thermal
motion only. The rapidity spectrum characterizes the system
development in the longitudinal direction and allows extracting
such an important feature of the system as the rapidity shift and
the width of the distribution. In the center-of-mass energy,
ranging from a few GeV to above ten TeV, the highest nowadays
collision energy available [1--4], the changing features of the
transverse momentum and rapidity spectra bring crucial information
about the changes in the particle production process.

To describe the transverse momentum and rapidity spectra, one can
use different distributions under the assumption of isotropic
emission in the transverse plane and consider rapidity shift in
the longitudinal direction. The distributions for transverse
momentum spectrum include, but are not limited to, the standard
(Fermi-Dirac, Bose-Einstein and Boltzmann) distribution [5--8],
Tsallis distribution [8--15], Erlang distribution [15], inverse
power-law [16--18], Schwinger mechanism [19--22], exponential
function and blast-wave model [23--26]. In some cases, one
distribution has different forms or revisions. The distributions
for rapidity spectrum includes mainly Gaussian [27--37],
double-Gaussian and multi-Gaussian functions [38--40], as well as
Gaussian-like functional forms. From a few GeV to above ten TeV,
it is expected that the interacting system undergoes different
phase states and has different reaction mechanisms. One can use
different distributions or the same distribution with different
parameter values to describe the different spectra.

Based on the descriptions of transverse momentum and rapidity
spectra, one can structure event patterns by using scatter plots
of identified particles at the stage of kinetic freeze-out of
interacting system. Our previous works [41--43] show that
different kinds of particles present different scatter plots thus
different event patterns. In particular, the event patterns
displayed by the scatter plots of light and heavy flavor particles
in three-dimensional velocity space are spherical and cylindrical
respectively, which show an obvious difference in the shape and
reflect different production mechanisms or stages. In addition, if
we consider the effects of non-zero elliptic flow on the two
spectra, more accurate event patterns can be obtained [44].
However, the effects of non-zero elliptic flow on the two spectra
and event patterns are in fact very small. One can neglect
non-zero elliptic flow in the case of describing the two spectra
and structuring the event patterns.

In most cases, the transverse momentum spectra are measured at
mid-rapidity or in the vicinity of a given rapidity only. These
spectra then often applied at other rapidities which leads to the
rapidity-independent spectra. These spectra have been used in our
previous works [41--44] leading to the event patterns being
non-fine in this sense. Here, we are interested in the use of the
rapidity-dependent spectra measured at different rapidities so
structuring the fine event patterns. For a not wide transverse
momentum spectrum (less than of $\sim 2$--3 GeV), one can use
either the standard distribution or the one combining two standard
distributions (``two-component standard distribution") with
different parameter values to describe the rapidity-dependent
spectra. In the distributions, the chemical potential and spin
property of considered particles can be included if the collision
energy per nucleon in center-of-mass system is not too high (less
than dozens of GeV). Generally, rapidity spectrum can be described
by Gaussian, double-Gaussian, or even three-Gaussian distribution
in our studies.

In this paper, rapidity-dependent double-differential transverse
momentum spectra of negatively charged pions ($\pi^-$) measured at
different rapidities and rapidity spectra of $\pi^-$ produced in
proton-proton (pp) collisions at the Super Proton Synchrotron
(SPS) at its Beam Energy Scan (BES) energies are investigated.
Related parameters are extracted by fitting the experimental data
of the NA61/SHINE Collaboration [45] and fine event patterns in pp
collisions are structured by fine scatter plots of $\pi^-$ due to
rapidity-dependent spectra being used. As a comparison, non-fine
results extracted from the experimental transverse mass spectra at
mid-rapidity and rapidity spectra of $\pi^-$ produced in pp
collisions and central (0--5\%) beryllium-beryllium (Be-Be),
argon-scandium (Ar-Sc), as well as lead-lead (Pb-Pb) collisions at
the SPS at its BES energies [46--50] are presented, where the
transverse mass spectra at mid-rapidity are used for whole
rapidity range. The transverse mass spectra used in the present
work are rapidity-independent.
\\

{\section{The model and method}}

As what we did in our recent work [44], first of all, we structure
a right-handed coordinate system and definite some variables
before introducing the model and method. Let the collision point
be the original $O$, one of the beam directions be the $Oz$ axis
and the reaction plane be the $xOz$ plane. Thus, the transverse
plane is the $xOy$ plane, the $Ox$ axis is along the impact
parameter and the $Oy$ axis is perpendicular to the $xOz$ plane.
Further, the velocity (momentum) components on the $Ox$, $Oy$ and
$Oz$ axes are denoted by $\beta_x$, $\beta_y$ and $\beta_z$
($p_x$, $p_y$ and $p_z$), respectively. According to rapidity $Y$
defined generally by energy $E$ and $p_z$, one can define rapidity
$Y_1$ by $E$ and $p_x$, as well as rapidity $Y_2$ by $E$ and
$p_y$. Finally, the three-dimensional velocity
($\beta_x$-$\beta_y$-$\beta_z$), momentum ($p_x$-$p_y$-$p_z$) and
rapidity ($Y_1$-$Y_2$-$Y$) spaces are structured by us.

The model used in the present work is in fact a hybrid model. In
consideration of different functions in descriptions of transverse
momentum ($p_{\rm T}$) or transverse mass ($m_{\rm T}$) spectra of
particles produced in soft process, we choose the standard
distribution and its superposition if necessary. In description of
$Y$ spectrum of particles, we choose a double-Gaussian function.
The center-of-mass energy range focused on in the present work is
from 6.1 (6.3) to 16.8 (17.3) GeV which is the SPS at its BES
energies. In this energy range, the chemical potential and spin
property of considered particles are included in our treatment on
soft excitation process, and the contribution of hard scattering
process can be neglected due to its nearly zero fraction.

The standard distribution has more than one forms. In the case of
considering $p_{\rm T}$ spectrum, we choose the probability
density function [8]
\begin{align}
f_{p_{\rm T}}(p_{\rm T},T) = & \frac{1}{N}\frac{dN}{dp_{\rm T}}=
Cp_{\rm T}m_{\rm T}\int_{Y_{\min}}^{Y_{\max}} \cosh Y \nonumber\\
& \times \bigg[\exp\bigg(\frac{m_{\rm T}\cosh Y-\mu}{T}
\bigg)+S\bigg]^{-1}dY,
\end{align}
where $N$ denotes the number of particles, $C$ denotes the
normalization constant, $m_{\rm T}=\sqrt{p_{\rm T}^2+m_0^2}$ and
$m_0$ is the rest mass, $Y_{\min}$ is the minimum $Y$ and
$Y_{\max}$ is the maximum $Y$ in the case of shifting the
mid-value of $Y$ to 0, $\mu$ is the chemical potential, $T$ is the
effective temperature and $S=+1$ and $-1$ correspond to fermions
and bosons respectively. In particular, $S=+1$, $-1$ and 0 also
correspond to Fermi-Dirac, Bose-Einstein and Boltzmann statistics,
respectively. In the case of considering $m_{\rm T}$ spectrum, we
have the normalized standard distribution
\begin{align}
f_{m_{\rm T}}(m_{\rm T},T) = & \frac{1}{N}\frac{dN}{dm_{\rm T}}=
Cm_{\rm T}^2\int_{Y_{\min}}^{Y_{\max}}\cosh Y \nonumber\\
& \times \bigg[\exp\bigg(\frac{m_{\rm T}\cosh Y-\mu}{T}
\bigg)+S\bigg]^{-1}dY.
\end{align}

Because of the similarity between expressions of $f_{p_{\rm
T}}(p_{\rm T},T)$ and $f_{m_{\rm T}}(m_{\rm T},T)$, one can use
directly parameters obtained from one function to another one. For
wide $p_{\rm T}$ ($m_{\rm T}$) spectrum, a single standard
distribution may not fit the data simultaneously in the
low-$p_{\rm T}$ and high-$p_{\rm T}$ regions. This would require a
combination of two or three standard distributions given the
spectrum width. In the case of using the two-component standard
distribution, we have to use two effective temperatures, $T_1$ and
$T_2$, for the first and second components respectively.
Meanwhile, one relative fraction, $k_1$, for the first component
is needed. The two-component standard distribution is written to
be
\begin{equation}
f_{p_{\rm T}}(p_{\rm T})=k_1f_{p_{\rm T}}(p_{\rm
T},T_1)+(1-k_1)f_{p_{\rm T}}(p_{\rm T},T_2)
\end{equation}
for $p_{\rm T}$ spectrum, or
\begin{equation}
f_{m_{\rm T}}(m_{\rm T})=k_1f_{m_{\rm T}}(m_{\rm
T},T_1)+(1-k_1)f_{m_{\rm T}}(m_{\rm T},T_2)
\end{equation}
for $m_{\rm T}$ spectrum. The effective temperature of interacting
system is then obtained by $T=k_1T_1+(1-k_1)T_2$. Generally,
two-component standard distribution is enough to describe particle
spectra in soft process. Three- or multi-standard distribution is
not necessary.

In the above functions, $\mu$ should not be regarded as a free
parameter due to its insensitivity to $p_{\rm T}$ or $m_{\rm T}$
spectrum. Instead, $\mu$ for particle type $i$, denoted by
$\mu_i$, can be obtained from the ratio of negatively to
positively charged particles. The following expression for $\mu_i$
\begin{equation}
\mu_i=-\frac{1}{2}T_{\rm ch} \ln(k_i),
\end{equation}
is applied, which is shown [51] to well reproduce the ratio of
negatively to positively charged particles ($k_i$) relating it
with the chemical freeze-out temperature $T_{\rm ch}$ within the
thermal model [52]. A relation between the yield ($n$) and the
mass ($m$) obtained in [53, 54], leads to the ratio of the first
to the second particles:
\begin{equation}
\frac{n_1}{n_2}=\frac{\exp(m_2/T_{\rm ch})+S_2}{\exp(m_1/T_{\rm
ch})+S_1},
\end{equation}
where $S_1(S_2)=\pm1$ denote fermions and bosons respectively. In
central nucleus-nucleus collisions, an empirical expression for
$T_{\rm ch}$ is applied [55--58]:
\begin{equation}
T_{\rm ch}=0.164\bigg\{1+\exp\bigg[2.60-\frac{\ln(\sqrt{s_{\rm
NN}})} {0.45}\bigg]\bigg\}^{-1},
\end{equation}
where $\sqrt{s_{\rm NN}}$ is the center-of-mass energy. Both the
units of $T_{\rm ch}$ and $\sqrt{s_{\rm NN}}$ in Eq. (7) are in
GeV, though $T_{\rm ch}$ and other temperatures appear in the
units of MeV in many cases.

For $Y$ spectrum, one can use the normalized symmetrical
double-Gaussian function as used by the NA61/SHINE [45] and NA49
Collaborations [47, 48]. That is,
\begin{align}
f_Y(Y) = & \frac{1}{2\sqrt{2\pi} \sigma_{Y}} \bigg\{ \exp
\bigg[-\frac{\big( Y+\delta Y\big)^2}{2\sigma_{Y}^2} \bigg] \nonumber\\
&+ \exp \bigg[-\frac{\big( Y-\delta Y\big)^2}{2\sigma_{Y}^2}
\bigg] \bigg\},
\end{align}
where $\delta Y$ and $\sigma_Y$ denote the rapidity shift and
distribution width respectively. In the case of considering
asymmetrical double-Gaussian function, we have to introduce the
relative fraction of the first (or second) component. Meanwhile,
the rapidity shifts and distribution widths for the two components
are separately different from each other. The present work focuses
on symmetrical or approximately symmetrical collisions. Eq. (8) is
an appropriate treatment that is acceptable for us.

The Monte Carlo distribution generating method is used below to
obtain $p_{\rm T}$ and $m_{\rm T}$ sampled according to the above
functions $f_{p_{\rm T}}(p_{\rm T})$ and $f_{m_{\rm T}}(m_{\rm
T})$. Let $R$ denotes the random number distributed evenly in
$[0,1]$. The values of $p_{\rm T}$ can be obtained by
\begin{equation}
\int_0^{p_{\rm T}}f_{p_{\rm T}}(p'_{\rm T})dp'_{\rm T} <R
<\int_0^{p_{\rm T}+\delta p_{\rm T}}f_{p_{\rm T}}(p'_{\rm T})
dp'_{\rm T},
\end{equation}
or, the values of $m_{\rm T}$ can be obtained by
\begin{equation}
\int_{m_0}^{m_{\rm T}}f_{m_{\rm T}}(m'_{\rm T})dm'_{\rm T} <R
<\int_{m_0}^{m_{\rm T}+\delta m_{\rm T}}f_{m_{\rm T}}(m'_{\rm T})
dm'_{\rm T},
\end{equation}
where $\delta p_{\rm T}$ and $\delta m_{\rm T}$ denote small
shifts relative to $p_{\rm T}$ and $m_{\rm T}$, respectively. Our
initial target is to obtain $p_{\rm T}$. In the case of obtaining
$m_{\rm T}$, we have $p_{\rm T}=\sqrt{m_{\rm T}^2-m_0^2}$.

Under the assumption of isotropic emission in the transverse
plane, we have
\begin{equation}
p_x=p_{\rm T} \cos \varphi=p_{\rm T} \cos(2\pi R_0)
\end{equation}
and
\begin{equation}
p_y=p_{\rm T}\sin \varphi =p_{\rm T} \sin(2\pi R_0)
\end{equation}
respectively, where
\begin{equation}
\varphi=\arctan \bigg (\frac{p_y}{p_x} \bigg)= 2\pi R_0
\end{equation}
denotes the azimuthal angle and $R_0$ denotes the random number
distributed evenly in $[0,1]$. Because of the assumption of
isotropic emission in the transverse plane, $p_x$ and $p_y$ do not
result in non-zero elliptic flow. As mentioned in the above
section, the effects of non-zero elliptic flow on $p_{\rm T}$
spectra, $Y$ spectra and scatter plots of particles are very small
and can be actually neglected [44]. Our assumption of isotropic
emission in the transverse plane is acceptable.

In the Monte Carlo method for the double-Gaussian function for $Y$
spectrum, let $R_{1,2}$ denote the random numbers distributed
evenly in $[0,1]$. We have
\begin{equation}
Y=\sigma_Y \sqrt{-2\ln R_1} \cos(2\pi R_2) -\delta Y
\end{equation}
in the calculation due to $Y$ expressed as above obeying Gaussian
function and the first component being Gaussian and similarly for
the second component. The $\cos(2\pi R_2)$ in the above expression
can be replaced by $\sin(2\pi R_2)$ [59]. In symmetrical
collisions, the two components have the same relative fraction,
i.e. 0.5. According to $Y$, we have
\begin{equation} p_z=m_{\rm T} \sinh Y
\end{equation}
and
\begin{equation}
E=\sqrt{p_z^2+m_{\rm T}^2}.
\end{equation}

Further,
\begin{equation}
\beta_{x}=\frac{p_{x}}{E},
\end{equation}
\begin{equation}
\beta_{y}=\frac{p_{y}}{E},
\end{equation}
\begin{equation}
\beta_{z}=\frac{p_{z}}{E},
\end{equation}
\begin{equation}
Y_1 \equiv \frac{1}{2}\ln \bigg( \frac{E+p_{x}}{E-p_{x}} \bigg),
\end{equation}
\begin{equation}
Y_2 \equiv \frac{1}{2}\ln \bigg( \frac{E+p_{y}}{E-p_{y}} \bigg).
\end{equation}
So far, the expressions of components in three-dimensional
$\beta_x$-$\beta_y$-$\beta_z$, $p_x$-$p_y$-$p_z$ and
$Y_1$-$Y_2$-$Y$ spaces are obtained.

The step-by-step calculations are made below, from Eq. (3) to Eqs.
(17--21), to obtain the event patterns in the three-dimensional
$\beta_x$-$\beta_y$-$\beta_z$, $p_x$-$p_y$-$p_z$ and
$Y_1$-$Y_2$-$Y$ plots.
\\

{\section{Results and discussion}}

\begin{figure*}[htb]
\begin{center}
\includegraphics[width=13.cm]{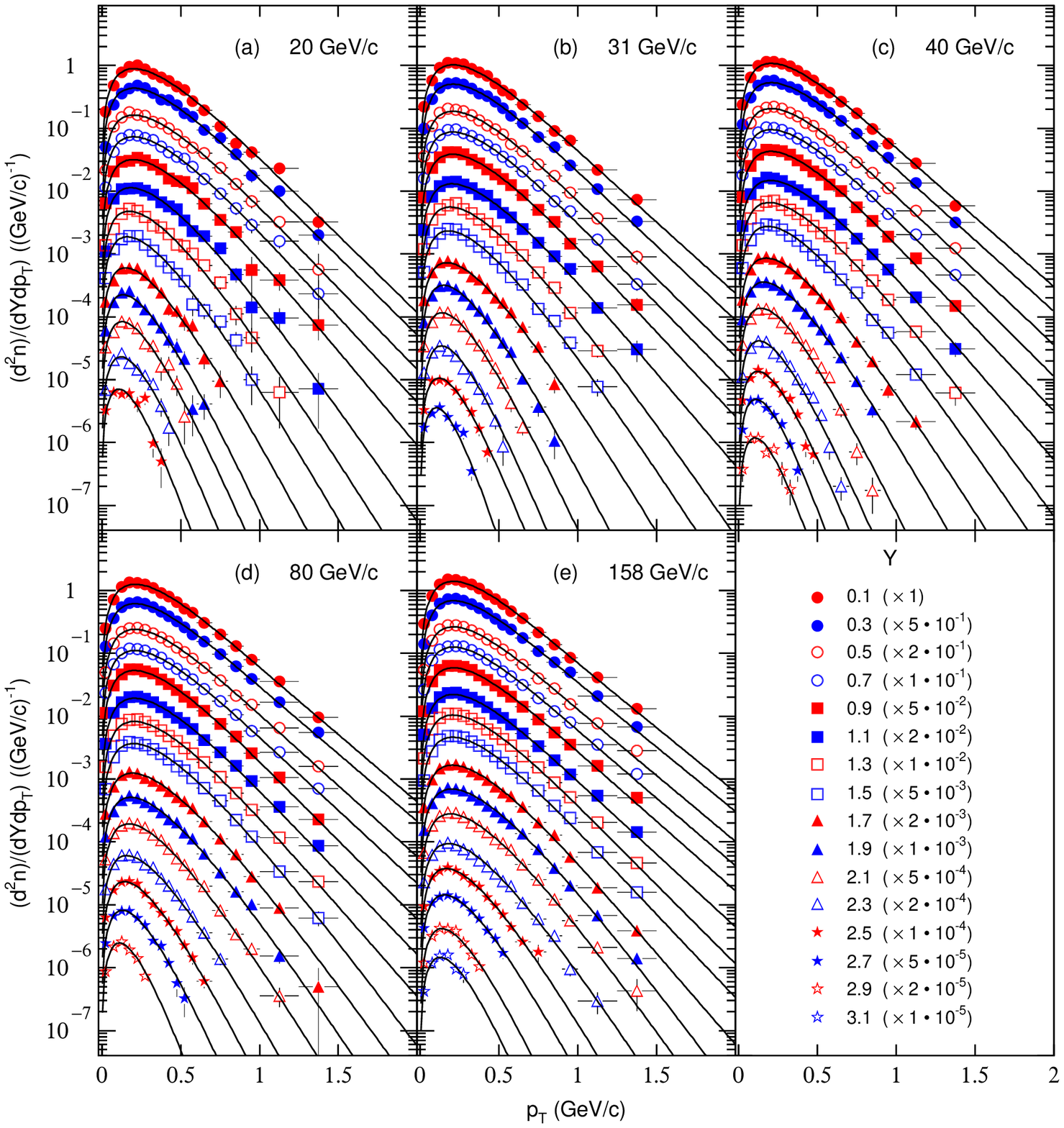}
\end{center}
\vskip-.0cm {\small Fig. 1. (color online) Rapidity-dependent
double-differential transverse momentum spectra of $\pi^-$ at
different rapidities $Y$ (rapidity bin centers) produced in pp
collisions at (a) 20, (b) 31, (c) 40, (d) 80 and (e) 158 GeV/$c$.
The symbols represent the experimental data of the NA61/SHINE
Collaboration [45] and the curves are the fits of (two-component)
standard distribution. The parameters are listed in Table 2.}
\end{figure*}

\begin{table*}
\vskip-1.0cm {\small Table 1. Values of $T_{\rm ch}$ and
$\mu_{\pi}$ obtained from Eqs. (5) and (6) (or (7)) in pp (or
central Pb-Pb) collisions at different $\sqrt{s_{\rm NN}}$ with
different $n_K/n_{\pi}$ (or $k_{\pi}$) read out from a table (or
plot) in ref. [60] (or [61]).
\begin{center}
\begin{tabular}{ccccc}
\hline\hline Type & $\sqrt{s_{\rm NN}}$/GeV & $T_{\rm ch}$/MeV & $\mu_{\pi}$/MeV & $n_K$/$n_{\pi}$ (or $k_{\pi}$)  \\
\hline
pp      & 6.3  & $126\pm3$ & $34.8\pm21.0$ & $0.574\pm0.205$ \\
        & 7.7  & $147\pm3$ & $25.3\pm19.3$ & $0.708\pm0.203$ \\
        & 8.8  & $153\pm4$ & $25.6\pm15.1$ & $0.716\pm0.149$ \\
        & 12.3 & $155\pm4$ & $21.3\pm12.6$ & $0.760\pm0.129$ \\
        & 17.3 & $160\pm4$ & $20.5\pm14.6$ & $0.773\pm0.155$ \\
\hline
Central & 6.3  & $134\pm3$ & $-10.3\pm4.8$ & $1.166\pm0.086$ \\
Pb-Pb   & 7.6  & $142\pm3$ & $-7.6\pm6.8$  & $1.112\pm0.140$ \\
        & 8.8  & $148\pm4$ & $-5.6\pm8.4$  & $1.092\pm0.147$ \\
        & 12.3 & $156\pm4$ & $-5.1\pm3.8$  & $1.067\pm0.074$ \\
        & 17.3 & $160\pm4$ & $-2.4\pm5.0$  & $1.031\pm0.086$ \\
\hline
\end{tabular}%
\end{center}}
\end{table*}

Figure 1 shows the rapidity-dependent $p_{\rm T}$ spectra,
$(d^2n)/(dY dp_{\rm T})$, of $\pi^-$ at different rapidities
produced in pp collisions at beam momentum $p_{\rm beam}=$ (a) 20,
(b) 31, (c) 40, (d) 80 and (e) 158 GeV/$c$ which correspond to
$\sqrt{s_{\rm NN}}=6.3$, 7.7, 8.8, 12.3 and 17.3 GeV,
respectively, where $n$ denote the number of $\pi^-$. The symbols
represent the experimental data of the NA61/SHINE Collaboration
[45] measured at different $Y$ and scaled by different amounts
marked in the last panel. The curves are the fits of standard (or
two-component standard) distribution. In the calculations, to
obtain correct values of parameters, we have to shift $Y$ to 0 and
use $Y_{\min}=-0.1$ and $Y_{\max}=0.1$ for each case. At different
$p_{\rm beam}$ and $\sqrt{s_{\rm NN}}$, the values of $\mu_{\pi}$
in pp collisions, based on Eqs. (5) and (6) and ref. [60] where
$n_{K}$ and $n_{\pi}$ are available to read out from a table, are
listed in Table 1 with $n_{K}/n_{\pi}$. The values of free
parameters ($T_1$, $T_2$ and $k_1$), derivative parameter ($T$),
normalization constant ($N_{p_{\rm T}}$), as well as $\chi^2$ and
degree of freedom (dof) in terms of $\chi^2$/dof are listed in
Table 2. One can see that the (two-component) standard
distribution describes well the NA61/SHINE experimental $p_{\rm
T}$ spectra of $\pi^-$ measured at different $Y$ in pp collisions
at the SPS at its BES energies. With increasing $Y$ and
$\sqrt{s_{\rm NN}}$, the parameters show some features discussed
below.

Figure 2 shows the $m_{\rm T}$ spectra, $1/m_{\rm T} \cdot
(d^2n)/(dY dm_{\rm T})$, of $\pi^-$ at mid-rapidity (0--0.2)
produced in pp collisions at $p_{\rm beam}=$ (a) 20, (b) 31, (c)
40, (d) 80 and (e) 158 GeV/$c$, in central (0--5\%) $^7$Be-$^9$Be
(Be-Be) collisions at (a) 20, (b) 30, (c) 40, (d) 75 and (e) 150
GeV/$c$, in central (0--5\%) Ar-Sc collisions at (a) 19, (b) 30,
(c) 40, (d) 75 and (e) 150 GeV/$c$, and in central (0--5\%) Pb-Pb
collisions at (a) 20, (b) 30, (c) 40, (d) 80 and (e) 158 GeV/$c$.
The values of $\sqrt{s_{\rm NN}}$ corresponding to $p_{\rm
beam}=19$, 30, 75 and 150 GeV/$c$ are 6.1, 7.6, 11.9 and 16.8 GeV,
respectively, where the relations for other $\sqrt{s_{\rm NN}}$
and $p_{\rm beam}$ as given just above and the asymmetries in
$^7$Be-$^9$Be and Ar-Sc collisions are neglected. The symbols
represent the experimental data of the NA61/SHINE [46] and NA49
Collaborations [47, 48]. The curves are the fits of
(two-component) standard distribution. The values of $\mu_{\pi}$
and $k_{\pi}$ in central Pb-Pb collisions, based on Eqs. (5) and
(7) and ref. [61] where $k_{\pi}$ is available to read out from a
plot, are listed in Table 1. The values of $\mu_{\pi}$ in central
Be-Be and Ar-Sc collisions are taken to be 0 due to $k_{\pi}$
being not available and $\mu_{\pi}$ having small effect on $\pi$
spectra. The values of $T_1$, $T_2$, $k_1$, $T$, normalization
constant ($N_{m_{\rm T}}$) and $\chi^2$/dof are listed in Table 3.
One can see that the (two-component) standard distribution
describes well the NA61/SHINE and NA49 experimental $m_{\rm T}$
spectra of $\pi^-$ measured in different collisions at the SPS at
its BES energies. With increasing system size ($A$) and
$\sqrt{s_{\rm NN}}$, where $A$ denotes the total nucleon number of
projectile and target nuclei, the parameters show some features
discussed below.

\begin{table*}
{\scriptsize Table 2. Values of free parameters ($T_1$, $T_2$ and
$k_1$), derivative parameter ($T$), normalization constant
($N_{p_{\rm T}}$) and $\chi^2$/dof for the curves fitted the
rapidity-dependent $p_{\rm T}$ spectra in Fig. 1.
\begin{center}
\begin{tabular}{cccccccccccc}
\hline\hline $\sqrt{s_{\rm NN}}$/GeV & $Y$ & $T_1$/MeV & $T_2$/MeV & $k_1$ & $T$/MeV & $N_{p_{\rm T}}$ & $\chi^2$/dof  \\
 \hline
6.3  & $0.0-0.2$ & $115\pm2$ & $154\pm3$ & $0.78\pm0.02$ & $124\pm2$ & $0.41\pm0.02$ & $12/14$ \\
     & $0.2-0.4$ & $115\pm2$ & $154\pm3$ & $0.78\pm0.02$ & $124\pm2$ & $0.40\pm0.02$ & $20/14$ \\
     & $0.4-0.6$ & $115\pm2$ & $150\pm3$ & $0.80\pm0.02$ & $122\pm2$ & $0.37\pm0.02$ & $12/14$ \\
     & $0.6-0.8$ & $115\pm2$ & $147\pm2$ & $0.85\pm0.02$ & $120\pm2$ & $0.33\pm0.02$ & $10/14$ \\
     & $0.8-1.0$ & $115\pm2$ & $140\pm2$ & $0.92\pm0.02$ & $117\pm2$ & $0.28\pm0.01$ & $11/14$ \\
     & $1.0-1.2$ & $107\pm2$ & $-$       & $1.00\pm0.00$ & $107\pm2$ & $0.23\pm0.01$ & $21/16$ \\
     & $1.2-1.4$ & $98\pm2$  & $-$       & $1.00\pm0.00$ & $98\pm2$  & $0.18\pm0.01$ & $17/15$ \\
     & $1.4-1.6$ & $90\pm2$  & $-$       & $1.00\pm0.00$ & $90\pm2$  & $0.13\pm0.01$ & $10/14$ \\
     & $1.6-1.8$ & $82\pm2$  & $-$       & $1.00\pm0.00$ & $82\pm2$  & $0.10\pm0.01$ & $8/12$ \\
     & $1.8-2.0$ & $71\pm1$  & $-$       & $1.00\pm0.00$ & $71\pm1$  & $0.06\pm0.01$ & $12/11$ \\
     & $2.0-2.2$ & $66\pm1$  & $-$       & $1.00\pm0.00$ & $66\pm1$  & $0.04\pm0.01$ & $6/9$ \\
     & $2.2-2.4$ & $60\pm1$  & $-$       & $1.00\pm0.00$ & $60\pm1$  & $0.03\pm0.01$ & $3/7$ \\
     & $2.4-2.6$ & $51\pm1$  & $-$       & $1.00\pm0.00$ & $51\pm1$  & $0.02\pm0.01$ & $8/6$ \\
\hline
7.7  & $0.0-0.2$ & $115\pm2$ & $164\pm3$ & $0.77\pm0.02$ & $126\pm2$ & $0.48\pm0.02$ & $16/14$ \\
     & $0.2-0.4$ & $115\pm2$ & $159\pm3$ & $0.77\pm0.02$ & $125\pm2$ & $0.47\pm0.02$ & $11/14$ \\
     & $0.4-0.6$ & $115\pm2$ & $154\pm3$ & $0.78\pm0.02$ & $124\pm2$ & $0.43\pm0.02$ & $16/14$ \\
     & $0.6-0.8$ & $115\pm2$ & $151\pm3$ & $0.80\pm0.02$ & $122\pm2$ & $0.40\pm0.02$ & $14/14$ \\
     & $0.8-1.0$ & $115\pm2$ & $147\pm2$ & $0.83\pm0.02$ & $120\pm2$ & $0.35\pm0.02$ & $10/14$ \\
     & $1.0-1.2$ & $115\pm2$ & $143\pm2$ & $0.92\pm0.02$ & $117\pm2$ & $0.29\pm0.01$ & $18/14$ \\
     & $1.2-1.4$ & $108\pm2$ & $-$       & $1.00\pm0.00$ & $108\pm2$ & $0.23\pm0.01$ & $15/15$ \\
     & $1.4-1.6$ & $101\pm2$ & $-$       & $1.00\pm0.00$ & $101\pm2$ & $0.18\pm0.01$ & $11/15$ \\
     & $1.6-1.8$ & $93\pm2$  & $-$       & $1.00\pm0.00$ & $93\pm2$  & $0.13\pm0.01$ & $12/13$ \\
     & $1.8-2.0$ & $79\pm2$  & $-$       & $1.00\pm0.00$ & $79\pm2$  & $0.10\pm0.01$ & $10/13$ \\
     & $2.0-2.2$ & $71\pm1$  & $-$       & $1.00\pm0.00$ & $71\pm1$  & $0.07\pm0.01$ & $8/11$ \\
     & $2.2-2.4$ & $62\pm1$  & $-$       & $1.00\pm0.00$ & $62\pm1$  & $0.04\pm0.01$ & $5/9$ \\
     & $2.4-2.6$ & $58\pm1$  & $-$       & $1.00\pm0.00$ & $58\pm1$  & $0.03\pm0.01$ & $3/7$ \\
     & $2.6-2.8$ & $46\pm1$  & $-$       & $1.00\pm0.00$ & $46\pm1$  & $0.01+0.01$   & $7/5$ \\
\hline
8.8  & $0.0-0.2$ & $115\pm2$ & $167\pm3$ & $0.76\pm0.02$ & $127\pm2$ & $0.51\pm0.03$ & $23/14$ \\
     & $0.2-0.4$ & $115\pm2$ & $163\pm3$ & $0.77\pm0.02$ & $126\pm2$ & $0.50\pm0.03$ & $19/14$ \\
     & $0.4-0.6$ & $115\pm2$ & $159\pm3$ & $0.77\pm0.02$ & $125\pm2$ & $0.48\pm0.02$ & $30/14$ \\
     & $0.6-0.8$ & $115\pm2$ & $153\pm3$ & $0.78\pm0.02$ & $123\pm2$ & $0.44\pm0.02$ & $24/14$ \\
     & $0.8-1.0$ & $115\pm2$ & $148\pm2$ & $0.82\pm0.02$ & $121\pm2$ & $0.39\pm0.02$ & $26/14$ \\
     & $1.0-1.2$ & $115\pm2$ & $140\pm2$ & $0.88\pm0.02$ & $118\pm2$ & $0.33\pm0.02$ & $24/14$ \\
     & $1.2-1.4$ & $112\pm2$ & $-$       & $1.00\pm0.00$ & $112\pm2$ & $0.28\pm0.01$ & $23/16$ \\
     & $1.4-1.6$ & $105\pm2$ & $-$       & $1.00\pm0.00$ & $105\pm2$ & $0.22\pm0.01$ & $23/15$ \\
     & $1.6-1.8$ & $97\pm2$  & $-$       & $1.00\pm0.00$ & $97\pm2$  & $0.16\pm0.01$ & $14/15$ \\
     & $1.8-2.0$ & $87\pm2$  & $-$       & $1.00\pm0.00$ & $87\pm2$  & $0.12\pm0.01$ & $8/13$ \\
     & $2.0-2.2$ & $74\pm1$  & $-$       & $1.00\pm0.00$ & $74\pm1$  & $0.08\pm0.01$ & $45/13$ \\
     & $2.2-2.4$ & $65\pm1$  & $-$       & $1.00\pm0.00$ & $65\pm1$  & $0.05\pm0.01$ & $42/11$ \\
     & $2.4-2.6$ & $60\pm1$  & $-$       & $1.00\pm0.00$ & $60\pm1$  & $0.03\pm0.01$ & $11/8$ \\
     & $2.6-2.8$ & $52\pm1$  & $-$       & $1.00\pm0.00$ & $52\pm1$  & $0.02\pm0.01$ & $7/6$ \\
     & $2.8-3.0$ & $48\pm1$  & $-$       & $1.00\pm0.00$ & $48\pm1$  & $0.01+0.01$   & $7/5$ \\
\hline
12.3 & $0.0-0.2$ & $115\pm2$ & $170\pm3$ & $0.75\pm0.02$ & $129\pm2$ & $0.60\pm0.03$ & $19/14$ \\
     & $0.2-0.4$ & $115\pm2$ & $170\pm3$ & $0.75\pm0.02$ & $129\pm2$ & $0.59\pm0.03$ & $15/14$ \\
     & $0.4-0.6$ & $115\pm2$ & $164\pm3$ & $0.76\pm0.02$ & $127\pm2$ & $0.57\pm0.03$ & $14/14$ \\
     & $0.6-0.8$ & $115\pm2$ & $161\pm3$ & $0.77\pm0.02$ & $126\pm2$ & $0.52\pm0.03$ & $12/14$ \\
     & $0.8-1.0$ & $115\pm2$ & $154\pm3$ & $0.81\pm0.02$ & $122\pm2$ & $0.49\pm0.02$ & $16/14$ \\
     & $1.0-1.2$ & $115\pm2$ & $150\pm2$ & $0.82\pm0.02$ & $121\pm2$ & $0.44\pm0.02$ & $21/14$ \\
     & $1.2-1.4$ & $115\pm2$ & $142\pm2$ & $0.87\pm0.02$ & $119\pm2$ & $0.37\pm0.02$ & $19/14$ \\
     & $1.4-1.6$ & $115\pm2$ & $-$       & $1.00\pm0.00$ & $115\pm2$ & $0.32\pm0.02$ & $21/16$ \\
     & $1.6-1.8$ & $110\pm2$ & $-$       & $1.00\pm0.00$ & $110\pm2$ & $0.26\pm0.01$ & $18/16$ \\
     & $1.8-2.0$ & $102\pm2$ & $-$       & $1.00\pm0.00$ & $102\pm2$ & $0.20\pm0.01$ & $18/15$ \\
     & $2.0-2.2$ & $94\pm2$  & $-$       & $1.00\pm0.00$ & $94\pm2$  & $0.14\pm0.01$ & $9/15$ \\
     & $2.2-2.4$ & $85\pm2$  & $-$       & $1.00\pm0.00$ & $85\pm2$  & $0.10\pm0.01$ & $7/12$ \\
     & $2.4-2.6$ & $75\pm1$  & $-$       & $1.00\pm0.00$ & $75\pm1$  & $0.07\pm0.01$ & $11/11$ \\
     & $2.6-2.8$ & $66\pm1$  & $-$       & $1.00\pm0.00$ & $66\pm1$  & $0.04\pm0.01$ & $6/9$ \\
     & $2.8-3.0$ & $53\pm1$  & $-$       & $1.00\pm0.00$ & $53\pm1$  & $0.03\pm0.01$ & $3/4$ \\
\hline
17.3 & $0.0-0.2$ & $115\pm2$ & $176\pm3$ & $0.76\pm0.02$ & $130\pm2$ & $0.67\pm0.03$ & $11/14$ \\
     & $0.2-0.4$ & $115\pm2$ & $176\pm3$ & $0.76\pm0.02$ & $130\pm2$ & $0.66\pm0.03$ & $14/14$ \\
     & $0.4-0.6$ & $115\pm2$ & $173\pm3$ & $0.76\pm0.02$ & $129\pm2$ & $0.63\pm0.03$ & $15/14$ \\
     & $0.6-0.8$ & $115\pm2$ & $170\pm3$ & $0.76\pm0.02$ & $128\pm2$ & $0.60\pm0.03$ & $12/14$ \\
     & $0.8-1.0$ & $115\pm2$ & $166\pm3$ & $0.75\pm0.02$ & $128\pm2$ & $0.56\pm0.03$ & $17/14$ \\
     & $1.0-1.2$ & $115\pm2$ & $162\pm3$ & $0.76\pm0.02$ & $126\pm2$ & $0.52\pm0.03$ & $9/14$ \\
     & $1.2-1.4$ & $115\pm2$ & $153\pm3$ & $0.80\pm0.02$ & $123\pm2$ & $0.48\pm0.02$ & $7/14 $ \\
     & $1.4-1.6$ & $115\pm2$ & $150\pm2$ & $0.85\pm0.02$ & $120\pm2$ & $0.42\pm0.02$ & $15/14$ \\
     & $1.6-1.8$ & $115\pm2$ & $144\pm2$ & $0.93\pm0.02$ & $117\pm2$ & $0.37\pm0.02$ & $13/14$ \\
     & $1.8-2.0$ & $114\pm2$ & $-$       & $1.00\pm0.00$ & $114\pm2$ & $0.30\pm0.02$ & $8/16$ \\
     & $2.0-2.2$ & $108\pm2$ & $-$       & $1.00\pm0.00$ & $108\pm2$ & $0.23\pm0.01$ & $10/16$ \\
     & $2.2-2.4$ & $100\pm2$ & $-$       & $1.00\pm0.00$ & $100\pm2$ & $0.18\pm0.01$ & $11/15$ \\
     & $2.4-2.6$ & $89\pm2$  & $-$       & $1.00\pm0.00$ & $89\pm2$  & $0.13\pm0.01$ & $6/12$ \\
     & $2.6-2.8$ & $81\pm2$  & $-$       & $1.00\pm0.00$ & $81\pm2$  & $0.09\pm0.01$ & $2/9$ \\
     & $2.8-3.0$ & $70\pm1$  & $-$       & $1.00\pm0.00$ & $70\pm1$  & $0.06\pm0.01$ & $9/6$ \\
     & $3.0-3.2$ & $63\pm1$  & $-$       & $1.00\pm0.00$ & $63\pm1$  & $0.04\pm0.01$ & $12/4$ \\
\hline
\end{tabular}%
\end{center}}
\end{table*}

\begin{figure*}[htbp]
\begin{center}
\includegraphics[width=13.cm]{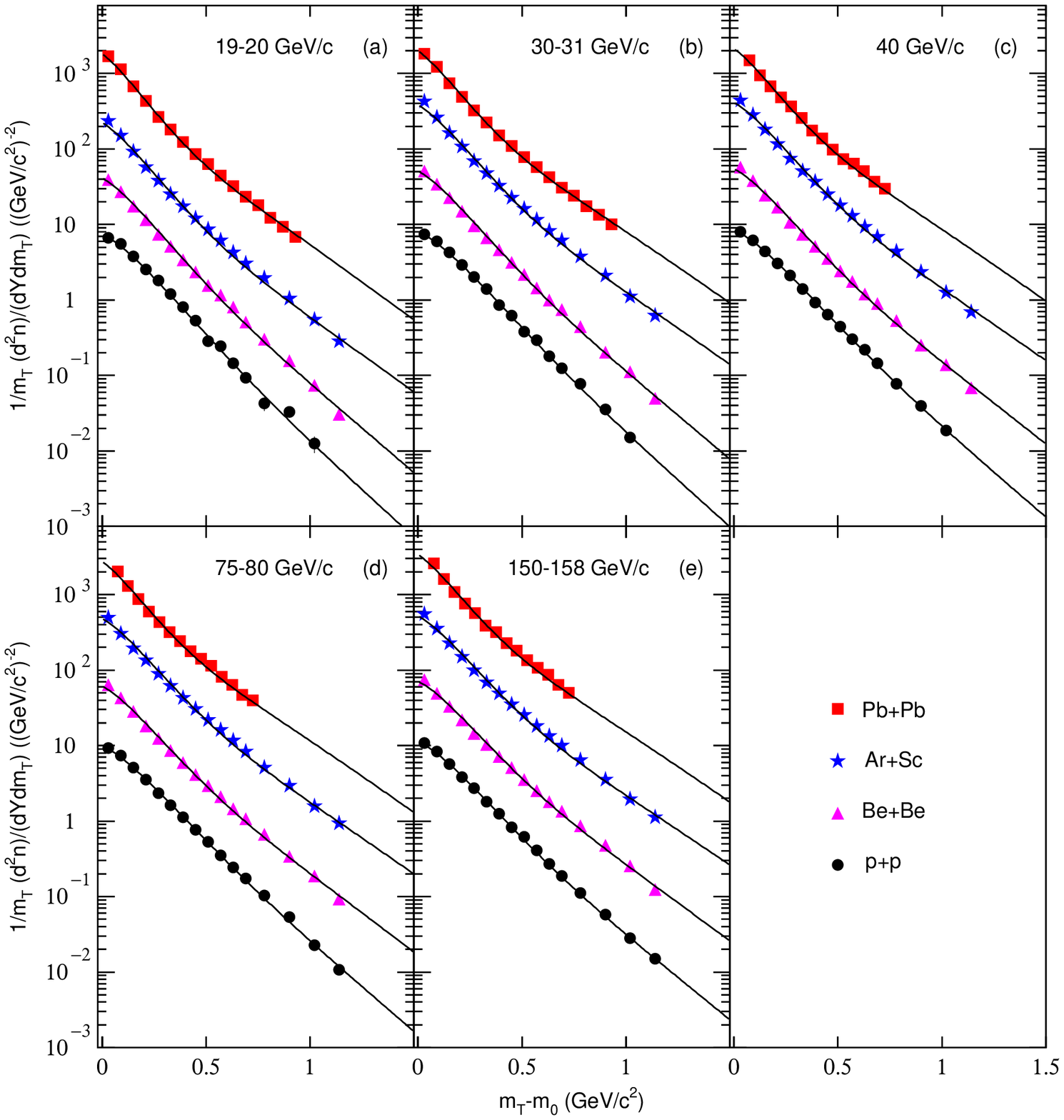}
\end{center}
\vskip-.0cm {\small Fig. 2. (color online) Double-differential
transverse mass spectra of $\pi^-$ at mid-rapidity (0--0.2)
produced in pp collisions at (a) 20, (b) 31, (c) 40, (d) 80 and
(e) 158 GeV/$c$ [46], in central Be-Be collisions at (a) 20, (b)
30, (c) 40, (d) 75 and (e) 150 GeV/$c$ [46], in central Ar-Sc
collisions at (a) 19, (b) 30, (c) 40, (d) 75 and (e) 150 GeV/$c$
[46], and in central Pb-Pb collisions at (a) 20, (b) 30, (c) 40,
(d) 80 and (e) 158 GeV/$c$ [47, 48]. The symbols represent the
experimental data of the NA61/SHINE [46] and NA49 Collaborations
[47, 48] and the curves are the fits of (two-component) standard
distribution. The parameters are listed in Table 3.}
\end{figure*}

\begin{table*}
\vskip-1.0cm {\scriptsize Table 3. Values of $T_1$, $T_2$, $k_1$,
$T$, normalization constant ($N_{m_{\rm T}}$) and $\chi^2$/dof for
the curves fitted the $m_{\rm T}$ spectra in Fig. 2.
\begin{center}
\begin{tabular}{ccccccccccccc}
\hline\hline Type & $\sqrt{s_{\rm NN}}$/GeV & $T_1$/MeV & $T_2$/MeV & $k_1$ & $T$/MeV & $N_{m_{\rm T}}$ & $\chi^2$/dof \\
\hline\
 pp    &  6.3 & $101\pm2$ & $161\pm3$ & $0.78\pm0.02$ & $114\pm2$ & $0.42\pm0.02$ & $10/10$ \\
       &  7.7 & $104\pm2$ & $164\pm3$ & $0.77\pm0.02$ & $118\pm2$ & $0.48\pm0.02$ & $4/10$ \\
       &  8.8 & $104\pm2$ & $169\pm3$ & $0.76\pm0.02$ & $120\pm2$ & $0.51\pm0.03$ & $3/10$ \\
       & 12.3 & $104\pm2$ & $169\pm3$ & $0.75\pm0.02$ & $120\pm2$ & $0.59\pm0.03$ & $4/11$ \\
       & 17.3 & $102\pm2$ & $177\pm3$ & $0.76\pm0.02$ & $120\pm2$ & $0.67\pm0.03$ & $3/11$ \\
\hline
Central&  6.3 &  $94\pm2$ & $167\pm3$ & $0.75\pm0.02$ & $112\pm2$ &  $2.1\pm0.1$  & $9/11$ \\
Be-Be  &  7.6 &  $95\pm2$ & $169\pm3$ & $0.73\pm0.02$ & $115\pm2$ &  $2.7\pm0.1$  & $10/11$ \\
       &  8.8 &  $97\pm2$ & $180\pm3$ & $0.74\pm0.02$ & $119\pm2$ &  $3.0\pm0.2$  & $7/11$ \\
       & 11.9 &  $96\pm2$ & $186\pm3$ & $0.72\pm0.02$ & $121\pm2$ &  $3.4\pm0.2$  & $7/11$ \\
       & 16.8 &  $98\pm2$ & $191\pm3$ & $0.72\pm0.02$ & $124\pm2$ &  $4.1\pm0.2$  & $10/11$ \\
\hline
Central&  6.1 &  $92\pm2$ & $193\pm3$ & $0.77\pm0.02$ & $115\pm2$ &   $11\pm1$    & $7/11$ \\
Ar-Sc  &  7.6 &  $94\pm2$ & $199\pm3$ & $0.75\pm0.02$ & $120\pm2$ &   $20\pm1$    & $8/11$ \\
       &  8.8 &  $95\pm2$ & $200\pm3$ & $0.75\pm0.02$ & $121\pm2$ &   $22\pm1$    & $12/11$ \\
       & 11.9 &  $95\pm2$ & $201\pm3$ & $0.74\pm0.02$ & $123\pm2$ &   $26\pm1$    & $9/11$ \\
       & 16.8 &  $96\pm2$ & $206\pm3$ & $0.73\pm0.02$ & $126\pm2$ &   $29\pm1$    & $8/11$ \\
\hline
Central&  6.3 &  $83\pm2$ & $190\pm3$ & $0.69\pm0.02$ & $116\pm2$ &   $81\pm4$    & $92/11$ \\
Pb-Pb  &  7.6 &  $85\pm2$ & $199\pm3$ & $0.66\pm0.02$ & $124\pm2$ &   $96\pm5$    & $70/11$ \\
       &  8.8 &  $84\pm2$ & $198\pm3$ & $0.65\pm0.02$ & $124\pm2$ &  $102\pm5$    & $79/9$ \\
       & 12.3 &  $84\pm2$ & $198\pm3$ & $0.64\pm0.02$ & $125\pm2$ &  $133\pm6$    & $102/9$ \\
       & 17.3 &  $86\pm2$ & $202\pm3$ & $0.65\pm0.02$ & $127\pm2$ &  $169\pm8$    & $93/9$ \\
\hline
\end{tabular}%
\end{center}}
\end{table*}

\begin{figure*}[htbp]
\begin{center}
\includegraphics[width=13.cm]{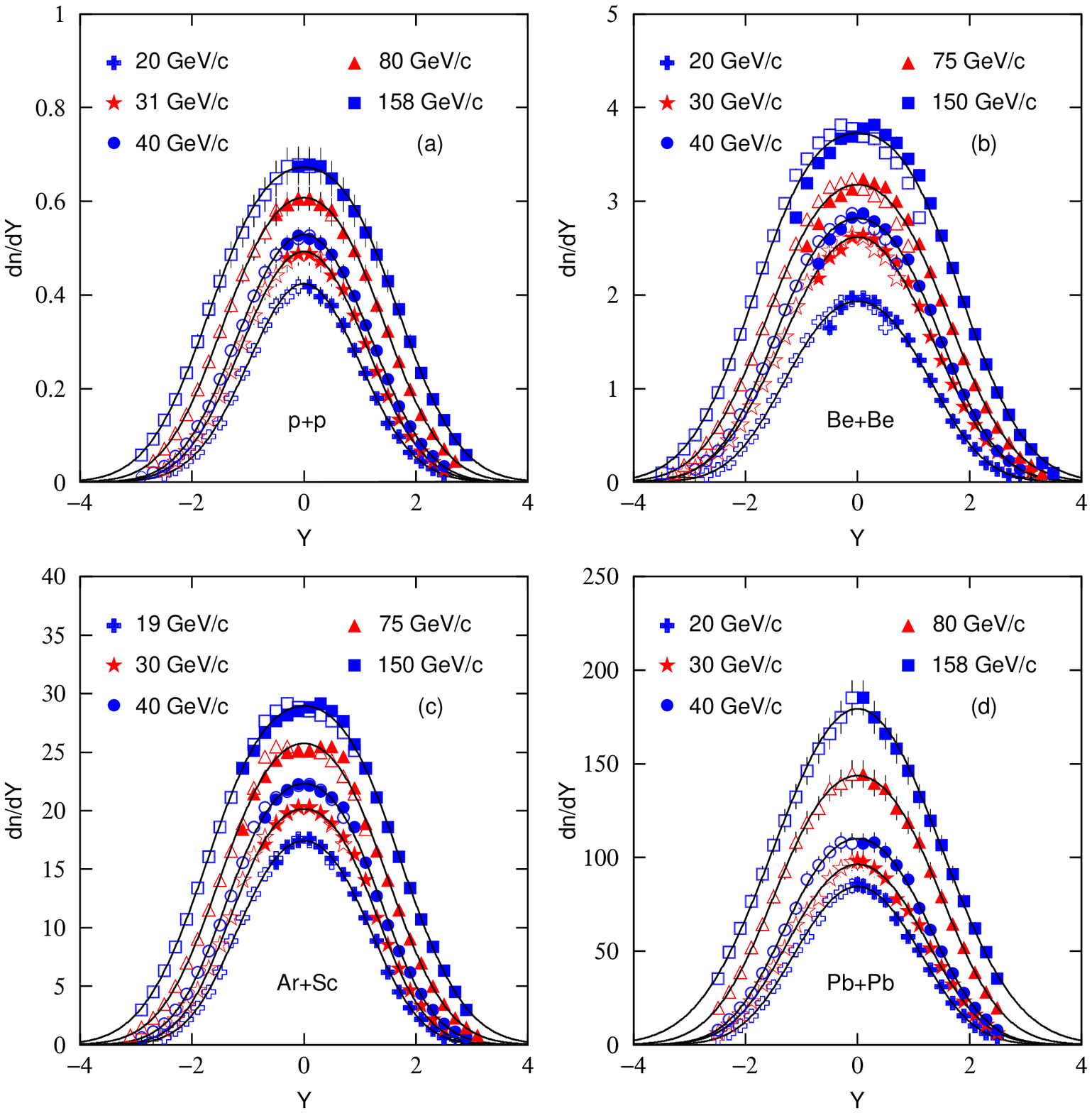}
\end{center}
\vskip-.0cm {\small Fig. 3. (color online) Rapidity spectra of
$\pi^-$ produced in (a) pp, (b) central Be-Be, (c) central Ar-Sc
[45, 49, 50] and (d) central Pb-Pb collisions [47, 48] at
different momenta marked in the panels. The closed symbols
represent the experimental data of the NA61/SHINE [45, 49, 50] and
NA49 Collaborations [47, 48] and the open ones are reflected at
$y=0$ [45, 47--50]. The curves are the fits of double-Gaussian
functions performed in refs. [45, 47, 48] for pp and central Pb-Pb
collisions or by us for central Be-Be and Ar-Sc collisions. The
fit parameters are given in Table 4.}
\end{figure*}

Figure 3 shows the $Y$ spectra, $dn/dY$, of $\pi^-$ produced in
(a) pp, (b) central Be-Be, (c) central Ar-Sc [45, 49, 50] and (d)
central Pb-Pb collisions [47, 48] at five momenta with little
variations marked in the panels. The values of $\sqrt{s_{\rm NN}}$
corresponding to different momenta are given in the just above.
The closed symbols represent the experimental data of the
NA61/SHINE [45, 49, 50] and NA49 Collaborations [47, 48] and the
open ones are reflected at $y=0$ [45, 47--50]. The curves are the
fits of double-Gaussian functions performed in refs. [45, 47, 48]
for pp and central Pb-Pb collisions or by us for central Be-Be and
Ar-Sc collisions. The values of free parameters ($\delta Y$ and
$\sigma_Y$) and normalization constant ($N_Y$) used in refs. [45,
47, 48] for pp and central Pb-Pb collisions or by us for central
Be-Be and Ar-Sc collisions, as well as $\chi^2/$dof are listed in
Table 4. One can see that the double-Gaussian function describes
well the NA61/SHINE and NA49 experimental $Y$ spectra of $\pi^-$
measured in different collisions at the SPS at its BES energies.
With increasing $A$ and $\sqrt{s_{\rm NN}}$, the parameters show
some features discussed below.

According to the parameter values obtained from Figs. 1 and 3 (or
similarly Figs. 2 and 3) and listed in Tables 2 and 4 (Tables 3
and 4), the Monte Carlo calculation can be performed and a series
of values of some kinematical quantities can be obtained.
According to these kinematical quantities, some scatter plots of
$\pi^-$ at the kinetic freeze-out can be structured. These scatter
plots are in fact the event patterns at the last stage of
collision process in the interacting system. Some characteristic
examples of the scatter plots are shown in Figs. 4 to 6.

\begin{table*}
\vskip-.0cm {\scriptsize Table 4. Values of free parameters
($\delta Y$ and $\sigma_Y$) and normalization constant ($N_Y$)
used in refs. [45, 47, 48] for pp and central Pb-Pb collisions or
by us for central Be-Be and Ar-Sc collisions, as well as
$\chi^2/$dof for the curves in Fig. 3.
\begin{center}
\begin{tabular}{cccccc}
\hline\hline Type & $\sqrt{s_{\rm NN}}$/GeV & $\delta Y$ & $\sigma_{Y}$ & $N_Y$ & $\chi^2$/dof \\
\hline
pp     & 6.3  & $0.337\pm0.046$ & $0.921\pm0.118$ & $1.05\pm0.05$ & $2/10$ \\
       & 7.7  & $0.545\pm0.055$ & $0.875\pm0.050$ & $1.31\pm0.07$ & $2/13$ \\
       & 8.8  & $0.604\pm0.044$ & $0.882\pm0.045$ & $1.48\pm0.05$ & $5/14$ \\
       & 12.3 & $0.733\pm0.010$ & $0.937\pm0.019$ & $1.94\pm0.08$ & $1/14$ \\
       & 17.3 & $0.860\pm0.021$ & $1.007\pm0.051$ & $2.44\pm0.13$ & $2/13$ \\
\hline
Central& 6.3  & $0.659\pm0.033$ & $0.875\pm0.044$ &  $5.6\pm0.05$ & $6/16$ \\
Be-Be  & 7.6  & $0.664\pm0.033$ & $1.006\pm0.050$ &  $8.2\pm0.07$ & $8/18$ \\
       & 8.8  & $0.753\pm0.038$ & $0.934\pm0.046$ &  $9.1\pm0.05$ & $10/18$ \\
       & 11.9 & $0.826\pm0.041$ & $1.026\pm0.050$ & $11.3\pm0.08$ & $27/20$ \\
       & 16.8 & $0.938\pm0.047$ & $1.080\pm0.050$ & $14.7\pm0.13$ & $59/21$ \\
\hline
Central& 6.1  & $0.605\pm0.030$ & $0.795\pm0.035$ & $46\pm2$ & $2/15$ \\
Ar-Sc  & 7.6  & $0.663\pm0.033$ & $0.810\pm0.040$ & $57\pm3$ & $6/16$ \\
       & 8.8  & $0.701\pm0.035$ & $0.859\pm0.042$ & $67\pm3$ & $7/16$ \\
       & 11.9 & $0.764\pm0.038$ & $0.953\pm0.047$ & $85\pm4$ & $23/19$ \\
       & 16.8 & $0.879\pm0.043$ & $1.042\pm0.050$ & $108\pm5$ & $8/18$ \\
\hline
Central& 6.3  & $0.557\pm0.009$ & $0.837\pm0.007$ &   $221\pm12$    & $2/11$ \\
Pb-Pb  & 7.6  & $0.624\pm0.009$ & $0.885\pm0.007$ &   $274\pm15$    & $3/11$ \\
       & 8.8  & $0.666\pm0.006$ & $0.872\pm0.005$ &   $322\pm19$    & $2/10$ \\
       & 12.3 & $0.756\pm0.006$ & $0.974\pm0.007$ &   $474\pm28$    & $1/10$ \\
       & 17.3 & $0.720\pm0.020$ & $1.180\pm0.020$ &   $639\pm48$    & $3/10$ \\
\hline
\end{tabular}%
\end{center}}
\end{table*}

Figure 4 shows a few scatter plots in pp and nucleus-nucleus
collisions in $\beta_x$-$\beta_y$-$\beta_z$ space. The values of
root-mean-squares $\sqrt{\overline{\beta_x^2}}$ for $\beta_x$,
$\sqrt{\overline{\beta_y^2}}$ for $\beta_y$ and
$\sqrt{\overline{\beta_z^2}}$ for $\beta_z$, as well as the
maximum $|\beta_x|$, $|\beta_y|$ and $|\beta_z|$ (i.e.
$|\beta_x|_{\max}$, $|\beta_y|_{\max}$ and $|\beta_z|_{\max}$) for
various systems and energies are listed in Table 5. One can see
that both the event patterns in $\beta_x$-$\beta_y$-$\beta_z$
space obtained from the rapidity-dependent and
rapidity-independent spectra in pp collisions are spherical,
though $\sqrt{\overline{\beta_{x,y}^2}}$
($\sqrt{\overline{\beta_z^2}}$) in fine event pattern is equal to
or larger (less) than that in non-fine event pattern. From pp to
central Pb-Pb collisions, there is no obvious change in the
non-fine event patterns. However, with increase of $\sqrt{s_{\rm
NN}}$, the size in transverse plane decreases and that in
longitudinal direction increases.

In Fig. 5, the scatter plots in $p_x$-$p_y$-$p_z$ space are shown
for the same collisions as in Fig. 4. The values of
root-mean-squares $\sqrt{\overline{p_x^2}}$ for $p_x$,
$\sqrt{\overline{p_y^2}}$ for $p_y$ and $\sqrt{\overline{p_z^2}}$
for $p_z$, as well as the maximum $|p_x|$, $|p_y|$ and $|p_z|$
(i.e. $|p_x|_{\max}$, $|p_y|_{\max}$ and $|p_z|_{\max}$) for
various systems and energies are listed in Table 6. One can see
that both the event patterns in $p_x$-$p_y$-$p_z$ space obtained
from the rapidity-dependent and rapidity-independent spectra in pp
collisions are rough cylindrical, though the size of fine event
pattern is smaller than that of non-fine event patterns. From pp
to central Pb-Pb collisions, there is no obvious change in the
size of non-fine event patterns. With increase of $\sqrt{s_{\rm
NN}}$, the size in transverse plane does not change obviously, and
that in longitudinal direction increases obviously.

\begin{figure*}[htbp]
\begin{center}
\includegraphics[width=13.cm]{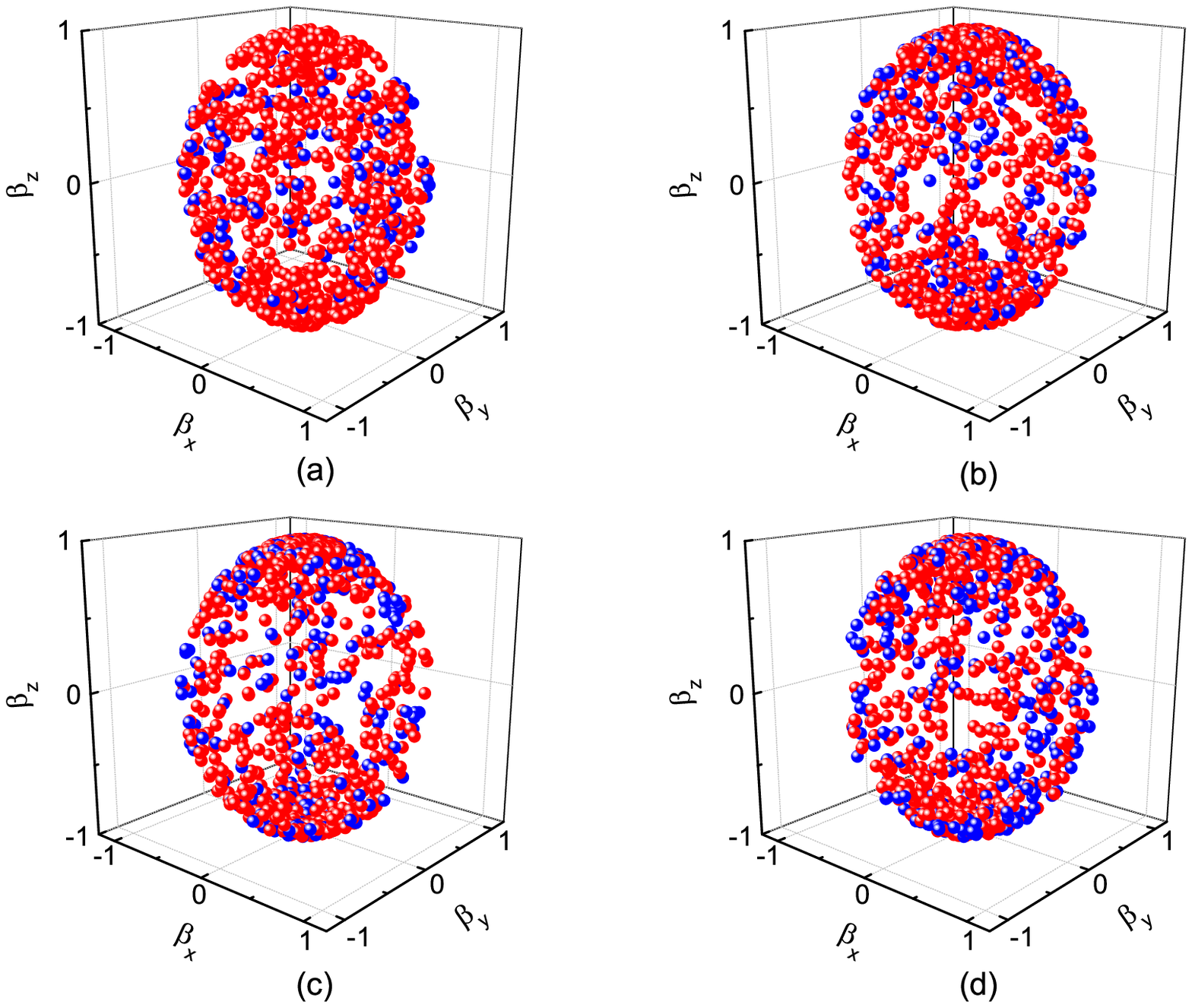}
\end{center}
\vskip-.0cm {\small Fig. 4. (color online) The scatter plots of
$\pi^-$ at $\sqrt{s_{\rm NN}}=6.3$ GeV in
$\beta_x$-$\beta_y$-$\beta_z$ space for (a) fine-event and (b)
non-fine scatter plots in pp collisions, and for non-fine scatter
plots for (c) Be-Be and (d) Pb-Pb collisions. The red and blue
globules represent the contributions of the first and second
components in the two-component standard distribution
respectively. The number of $\pi^-$ for each panel is 1000.}
\end{figure*}

\begin{table*}
\vskip-.0cm {\scriptsize Table 5. Values of the root-mean-squares
$\sqrt{\overline{\beta_x^2}}$ for $\beta_x$,
$\sqrt{\overline{\beta_y^2}}$ for $\beta_y$ and
$\sqrt{\overline{\beta_z^2}}$ for $\beta_z$, and the maximum
$|\beta_x|$, $|\beta_y|$ and $|\beta_z|$ ($|\beta_x|_{\max}$,
$|\beta_y|_{\max}$ and $|\beta_z|_{\max}$) corresponding to the
scatter plots for different collisions (see examples in Fig. 4).
Both the root-mean-squares and the maximum velocity components are
in the units of $c$.
\begin{center}
\begin{tabular}{ccccccccc}
\hline\hline Type & $\sqrt{s_{\rm NN}}$/GeV & $\sqrt{\overline{\beta_x^2}}$ & $\sqrt{\overline{\beta_y^2}}$ & $\sqrt{\overline{\beta_z^2}}$ & $|\beta_x|_{\max}$ & $|\beta_y|_{\max}$ & $|\beta_z|_{\max}$ \\
\hline
Fine     & 6.3  & $0.461\pm0.009$ & $0.454\pm0.008$ & $0.611\pm0.008$ & $0.984$ & $0.965$ & $0.989$ \\
pp     & 7.7  & $0.439\pm0.009$ & $0.433\pm0.009$ & $0.636\pm0.008$ & $0.990$ & $0.966$ & $0.991$ \\
         & 8.8  & $0.431\pm0.009$ & $0.426\pm0.009$ & $0.650\pm0.008$ & $0.964$ & $0.975$ & $0.995$ \\
         & 12.3 & $0.402\pm0.010$ & $0.395\pm0.009$ & $0.684\pm0.008$ & $0.974$ & $0.979$ & $0.994$ \\
         & 17.3 & $0.372\pm0.010$ & $0.371\pm0.010$ & $0.716\pm0.007$ & $0.982$ & $0.967$ & $0.997$ \\
\hline
Non-fine & 6.3  & $0.425\pm0.008$ & $0.418\pm0.008$ & $0.727\pm0.007$ & $0.972$ & $0.985$ & $0.998$ \\
pp     & 7.7  & $0.418\pm0.008$ & $0.407\pm0.008$ & $0.743\pm0.007$ & $0.974$ & $0.976$ & $0.998$ \\
         & 8.8  & $0.409\pm0.008$ & $0.405\pm0.008$ & $0.754\pm0.007$ & $0.970$ & $0.991$ & $0.999$ \\
         & 12.3 & $0.391\pm0.008$ & $0.384\pm0.008$ & $0.781\pm0.007$ & $0.975$ & $0.970$ & $0.998$ \\
         & 17.3 & $0.371\pm0.009$ & $0.365\pm0.008$ & $0.804\pm0.007$ & $0.975$ & $0.964$ & $0.999$ \\
\hline
Central  & 6.3  & $0.401\pm0.008$ & $0.390\pm0.008$ & $0.765\pm0.007$ & $0.943$ & $0.954$ & $0.997$ \\
Be-Be    & 7.7  & $0.393\pm0.008$ & $0.381\pm0.008$ & $0.778\pm0.007$ & $0.961$ & $0.966$ & $0.998$ \\
         & 8.8  & $0.385\pm0.008$ & $0.378\pm0.008$ & $0.785\pm0.007$ & $0.973$ & $0.976$ & $0.998$ \\
         & 12.3 & $0.379\pm0.009$ & $0.366\pm0.009$ & $0.801\pm0.007$ & $0.963$ & $0.978$ & $0.999$ \\
         & 17.3 & $0.366\pm0.009$ & $0.353\pm0.009$ & $0.818\pm0.006$ & $0.968$ & $0.975$ & $0.999$ \\
\hline
Central  & 6.3  & $0.421\pm0.008$ & $0.405\pm0.008$ & $0.744\pm0.007$ & $0.973$ & $0.984$ & $0.997$ \\
Ar-Sc    & 7.7  & $0.412\pm0.008$ & $0.399\pm0.008$ & $0.756\pm0.007$ & $0.960$ & $0.983$ & $0.998$ \\
         & 8.8  & $0.403\pm0.008$ & $0.391\pm0.008$ & $0.768\pm0.007$ & $0.974$ & $0.972$ & $0.999$ \\
         & 12.3 & $0.385\pm0.008$ & $0.384\pm0.008$ & $0.787\pm0.007$ & $0.963$ & $0.983$ & $0.999$ \\
         & 17.3 & $0.368\pm0.009$ & $0.367\pm0.009$ & $0.807\pm0.007$ & $0.991$ & $0.978$ & $0.999$ \\
\hline
Central  & 6.3  & $0.417\pm0.008$ & $0.409\pm0.008$ & $0.741\pm0.007$ & $0.969$ & $0.984$ & $0.998$ \\
Pb-Pb    & 7.6  & $0.407\pm0.008$ & $0.400\pm0.008$ & $0.757\pm0.007$ & $0.954$ & $0.981$ & $0.998$ \\
         & 8.8  & $0.404\pm0.008$ & $0.395\pm0.008$ & $0.763\pm0.007$ & $0.951$ & $0.995$ & $0.999$ \\
         & 12.3 & $0.391\pm0.008$ & $0.384\pm0.008$ & $0.781\pm0.007$ & $0.975$ & $0.970$ & $0.998$ \\
         & 17.3 & $0.378\pm0.009$ & $0.364\pm0.009$ & $0.798\pm0.007$ & $0.961$ & $0.981$ & $0.999$ \\
\hline
\end{tabular}%
\end{center}}
\end{table*}

\begin{figure*}[htbp]
\begin{center}
\includegraphics[width=13.cm]{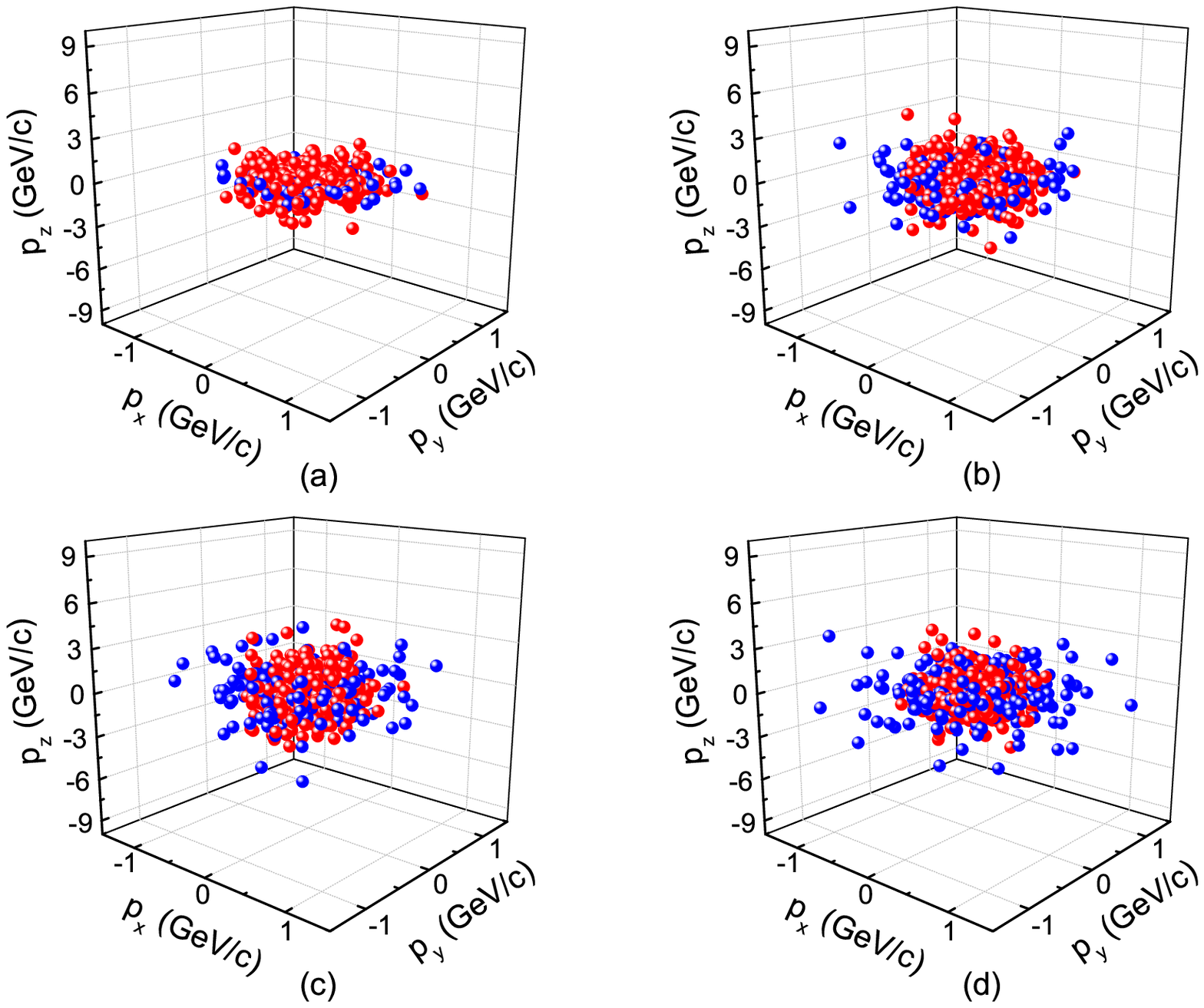}
\end{center}
\vskip-.0cm {\small Fig. 5. (color online) Same as Fig. 4, but for
$p_x$-$p_y$-$p_z$ space.}
\end{figure*}

\begin{table*}
\vskip-.0cm {\scriptsize Table 6. Values of the root-mean-squares
$\sqrt{\overline{p_x^2}}$ for $p_x$, $\sqrt{\overline{p_y^2}}$ for
$p_y$ and $\sqrt{\overline{p_z^2}}$ for $p_z$, and the maximum
$|p_x|$, $|p_y|$ and $|p_z|$ ($|p_x|_{\max}$, $|p_y|_{\max}$ and
$|p_z|_{\max}$) corresponding to the scatter plots for different
collisions (see examples in Fig. 5). All the root-mean-squares and
the maximum momentum components are in the units of GeV/$c$.
\begin{center}
\begin{tabular}{ccccccccc}
\hline\hline  Type & $\sqrt{s_{\rm NN}}$/GeV & $\sqrt{\overline{p_x^2}}$ & $\sqrt{\overline{p_y^2}}$ & $\sqrt{\overline{p_z^2}}$ & $|p_x|_{\max}$ & $|p_y|_{\max}$ & $|p_z|_{\max}$ \\
\hline
Fine     & 6.3  & $0.262\pm0.008$ & $0.242\pm0.007$ & $0.437\pm0.017$ & $1.195$ & $0.923$ & $2.359$ \\
pp     & 7.7  & $0.254\pm0.008$ & $0.233\pm0.007$ & $0.483\pm0.019$ & $1.275$ & $0.925$ & $2.573$ \\
         & 8.8  & $0.258\pm0.008$ & $0.240\pm0.007$ & $0.535\pm0.021$ & $1.235$ & $0.934$ & $2.653$ \\
         & 12.3 & $0.252\pm0.009$ & $0.231\pm0.007$ & $0.653\pm0.028$ & $1.315$ & $0.964$ & $3.543$ \\
         & 17.3 & $0.232\pm0.008$ & $0.226\pm0.009$ & $0.791\pm0.038$ & $1.055$ & $1.364$ & $4.470$ \\
\hline
Non-fine & 6.3  & $0.288\pm0.008$ & $0.278\pm0.008$ & $0.878\pm0.033$ & $1.329$ & $1.100$ & $5.037$ \\
pp     & 7.7  & $0.298\pm0.009$ & $0.285\pm0.008$ & $0.982\pm0.037$ & $1.348$ & $1.118$ & $5.659$ \\
         & 8.8  & $0.302\pm0.009$ & $0.292\pm0.009$ & $1.095\pm0.054$ & $1.367$ & $1.356$ & $8.376$ \\
         & 12.3 & $0.303\pm0.009$ & $0.292\pm0.009$ & $1.276\pm0.050$ & $1.367$ & $1.356$ & $6.306$ \\
         & 17.3 & $0.304\pm0.009$ & $0.293\pm0.009$ & $1.531\pm0.056$ & $1.396$ & $1.397$ & $7.439$ \\
\hline
Central  & 6.3  & $0.286\pm0.009$ & $0.276\pm0.009$ & $0.997\pm0.032$ & $1.358$ & $1.346$ & $4.115$ \\
Be-Be    & 7.7  & $0.292\pm0.009$ & $0.283\pm0.009$ & $1.215\pm0.045$ & $1.367$ & $1.356$ & $6.338$ \\
         & 8.8  & $0.302\pm0.009$ & $0.294\pm0.009$ & $1.284\pm0.050$ & $1.405$ & $1.407$ & $6.414$ \\
         & 12.3 & $0.309\pm0.010$ & $0.299\pm0.010$ & $1.487\pm0.056$ & $1.425$ & $1.437$ & $7.794$ \\
         & 17.3 & $0.315\pm0.010$ & $0.305\pm0.010$ & $1.809\pm0.072$ & $1.444$ & $1.447$ & $9.762$ \\
\hline
Central  & 6.3  & $0.308\pm0.009$ & $0.299\pm0.010$ & $0.911\pm0.031$ & $1.313$ & $1.471$ & $4.078$ \\
Ar-Sc    & 7.7  & $0.312\pm0.010$ & $0.311\pm0.010$ & $1.012\pm0.037$ & $1.226$ & $1.422$ & $5.120$ \\
         & 8.8  & $0.313\pm0.010$ & $0.297\pm0.009$ & $1.113\pm0.042$ & $1.437$ & $1.324$ & $5.687$ \\
         & 12.3 & $0.299\pm0.009$ & $0.311\pm0.010$ & $1.312\pm0.052$ & $1.157$ & $1.378$ & $8.056$ \\
         & 17.3 & $0.311\pm0.009$ & $0.320\pm0.010$ & $1.683\pm0.070$ & $1.264$ & $1.397$ & $9.010$ \\
\hline
Central  & 6.3  & $0.309\pm0.009$ & $0.312\pm0.011$ & $0.918\pm0.033$ & $1.421$ & $1.559$ & $4.422$ \\
Pb-Pb    & 7.6  & $0.312\pm0.010$ & $0.319\pm0.011$ & $1.091\pm0.042$ & $1.226$ & $1.422$ & $5.299$ \\
         & 8.8  & $0.326\pm0.010$ & $0.331\pm0.012$ & $1.156\pm0.054$ & $1.546$ & $1.569$ & $9.169$ \\
         & 12.3 & $0.303\pm0.009$ & $0.292\pm0.009$ & $1.276\pm0.050$ & $1.367$ & $1.356$ & $6.306$ \\
         & 17.3 & $0.312\pm0.009$ & $0.314\pm0.011$ & $1.684\pm0.067$ & $1.290$ & $1.522$ & $8.891$ \\
\hline
\end{tabular}%
\end{center}}
\end{table*}

\begin{figure*}[htbp]
\begin{center}
\includegraphics[width=13.cm]{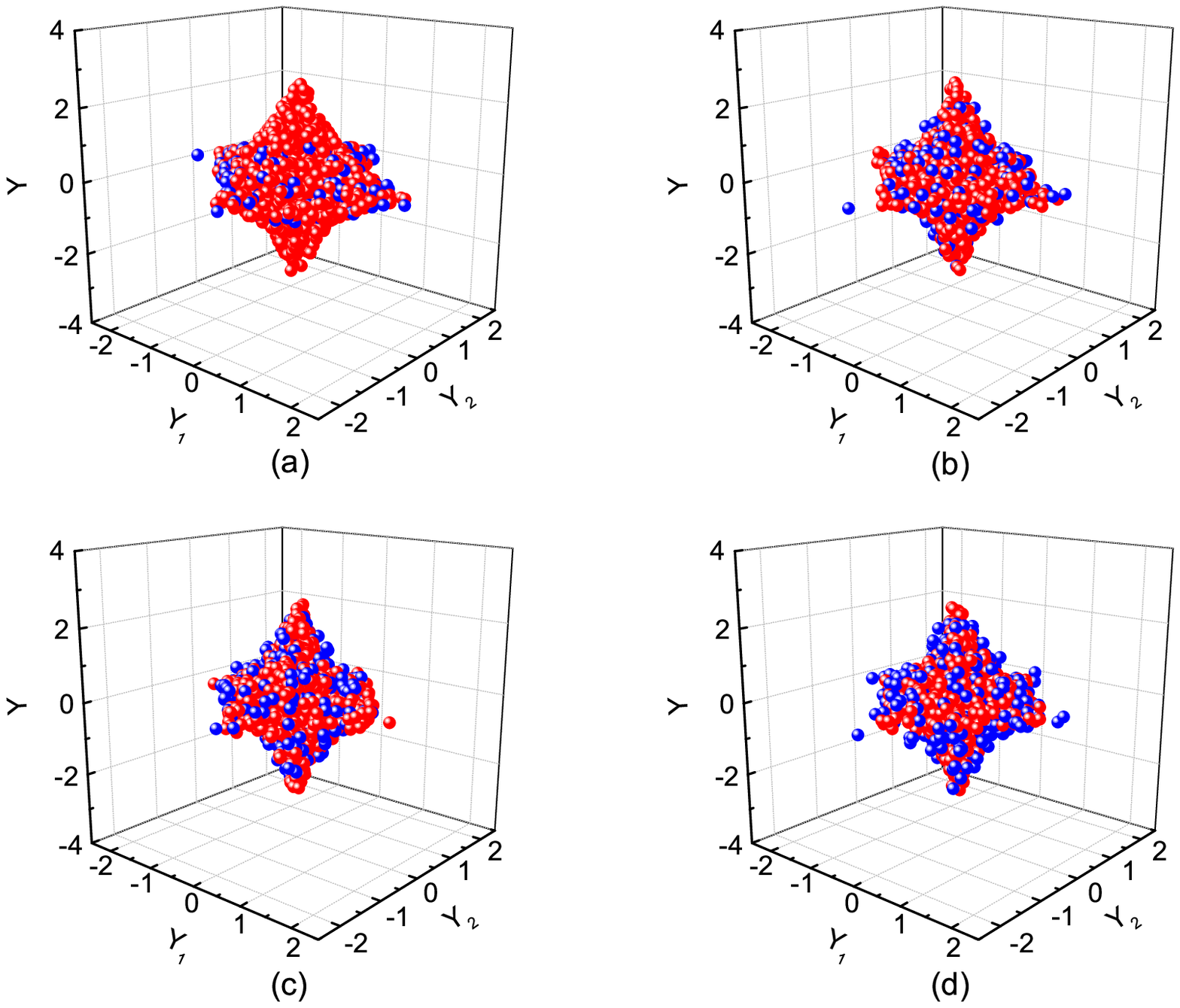}
\end{center}
\vskip-.0cm  {\small Fig. 6. (color online) Same as Fig. 4, but
for $Y_1$-$Y_2$-$Y$ space.}
\end{figure*}

\begin{table*}
\vskip-.0cm {\scriptsize Table 7. Values of the root-mean-squares
$\sqrt{\overline{Y_1^2}}$ for $Y_1$, $\sqrt{\overline{Y_2^2}}$ for
$Y_2$ and $\sqrt{\overline{Y^2}}$ for $Y$, and the maximum
$|Y_1|$, $|Y_2|$ and $|Y|$ ($|Y_1|_{\max}$, $|Y_2|_{\max}$ and
$|Y|_{\max}$) corresponding to the scatter plots for different
collisions (see examples in Fig. 6).
\begin{center}
\begin{tabular}{ccccccccc}
\hline\hline Type & $\sqrt{s_{\rm NN}}$/GeV & $\sqrt{\overline{Y_1^2}}$ & $\sqrt{\overline{Y_2^2}}$ &$\sqrt{\overline{Y^2}}$ & $|Y_1|_{\max}$ & $|Y_2|_{\max}$ & $|Y|_{\max}$ \\
\hline
Fine     & 6.3  & $0.626\pm0.017$ & $0.605\pm0.016$ & $0.938\pm0.020$ & $2.404$ & $2.014$ & $2.593$ \\
pp     & 7.7  & $0.602\pm0.018$ & $0.580\pm0.016$ & $1.001\pm0.020$ & $2.657$ & $2.022$ & $2.688$ \\
         & 8.8  & $0.579\pm0.016$ & $0.570\pm0.016$ & $1.049\pm0.021$ & $2.000$ & $2.180$ & $2.969$ \\
         & 12.3 & $0.553\pm0.018$ & $0.533\pm0.017$ & $1.158\pm0.022$ & $2.168$ & $2.285$ & $2.922$ \\
         & 17.3 & $0.507\pm0.008$ & $0.500\pm0.017$ & $1.285\pm0.024$ & $2.364$ & $2.049$ & $3.179$ \\
\hline
Non-fine & 6.3  & $0.547\pm0.015$ & $0.533\pm0.015$ & $0.938\pm0.020$ & $2.120$ & $2.448$ & $2.593$ \\
pp     & 7.7  & $0.537\pm0.015$ & $0.521\pm0.015$ & $1.001\pm0.020$ & $2.169$ & $2.208$ & $2.688$ \\
         & 8.8  & $0.528\pm0.016$ & $0.520\pm0.016$ & $1.049\pm0.021$ & $2.098$ & $2.708$ & $2.969$ \\
         & 12.3 & $0.504\pm0.016$ & $0.487\pm0.015$ & $1.158\pm0.022$ & $2.183$ & $2.091$ & $2.922$ \\
         & 17.3 & $0.478\pm0.016$ & $0.460\pm0.015$ & $1.285\pm0.024$ & $2.185$ & $2.003$ & $3.179$ \\
\hline
Central  & 6.3  & $0.510\pm0.014$ & $0.493\pm0.015$ & $1.052\pm0.020$ & $1.761$ & $1.879$ & $2.558$ \\
Be-Be    & 7.7  & $0.509\pm0.016$ & $0.487\pm0.015$ & $1.147\pm0.022$ & $1.953$ & $2.022$ & $2.766$ \\
         & 8.8  & $0.494\pm0.016$ & $0.479\pm0.015$ & $1.169\pm0.022$ & $2.137$ & $2.206$ & $2.935$ \\
         & 12.3 & $0.493\pm0.016$ & $0.472\pm0.016$ & $1.263\pm0.023$ & $1.990$ & $2.250$ & $2.970$ \\
         & 17.3 & $0.477\pm0.017$ & $0.457\pm0.017$ & $1.376\pm0.025$ & $2.053$ & $2.185$ & $3.194$ \\
\hline
Central  & 6.3  & $0.542\pm0.015$ & $0.516\pm0.015$ & $0.977\pm0.019$ & $2.147$ & $2.408$ & $2.552$ \\
Ar-Sc    & 7.7  & $0.527\pm0.015$ & $0.512\pm0.016$ & $1.026\pm0.020$ & $1.945$ & $2.382$ & $2.647$ \\
         & 8.8  & $0.516\pm0.015$ & $0.500\pm0.015$ & $1.090\pm0.021$ & $2.173$ & $2.122$ & $2.993$ \\
         & 12.3 & $0.494\pm0.015$ & $0.499\pm0.017$ & $1.190\pm0.022$ & $1.985$ & $2.366$ & $2.991$ \\
         & 17.3 & $0.474\pm0.017$ & $0.474\pm0.016$ & $1.315\pm0.024$ & $2.708$ & $2.259$ & $3.140$ \\
\hline
Central  & 6.3  & $0.536\pm0.015$ & $0.529\pm0.016$ & $0.975\pm0.019$ & $2.072$ & $2.420$ & $2.587$ \\
Pb-Pb    & 7.6  & $0.517\pm0.014$ & $0.519\pm0.016$ & $1.056\pm0.021$ & $1.879$ & $2.320$ & $2.792$ \\
         & 8.8  & $0.516\pm0.015$ & $0.525\pm0.019$ & $1.077\pm0.021$ & $1.840$ & $2.960$ & $2.993$ \\
         & 12.3 & $0.504\pm0.016$ & $0.487\pm0.015$ & $1.158\pm0.022$ & $2.183$ & $2.091$ & $2.922$ \\
         & 17.3 & $0.486\pm0.016$ & $0.465\pm0.016$ & $1.312\pm0.025$ & $1.952$ & $2.315$ & $3.185$ \\
\hline
\end{tabular}%
\end{center}}
\end{table*}

The scatter plots in $Y_1$-$Y_2$-$Y$ space are shown in Fig. 6.
The values of root-mean-squares $\sqrt{\overline{Y_1^2}}$ for
$Y_1$, $\sqrt{\overline{Y_2^2}}$ for $Y_2$ and
$\sqrt{\overline{Y^2}}$ for $Y$, as well as the maximum $|Y_1|$,
$|Y_2|$ and $|Y|$ (i.e. $|Y_1|_{\max}$, $|Y_2|_{\max}$ and
$|Y|_{\max}$) for various systems and energies are listed in Table
7. One can see that both the event patterns in $Y_1$-$Y_2$-$Y$
space obtained from the rapidity-dependent and
rapidity-independent spectra in pp collisions are rough
rhombohedral, though the size of fine event pattern in transverse
plane is larger than that of non-fine event patterns. However,
from pp to central Pb-Pb collisions, there is no obvious change in
the size of non-fine event patterns. With increase of
$\sqrt{s_{\rm NN}}$, the size in transverse plane decreases, and
that in longitudinal direction increases.

\begin{figure*}[htb]
\vskip-.0cm
\begin{center}
\includegraphics[width=13.cm]{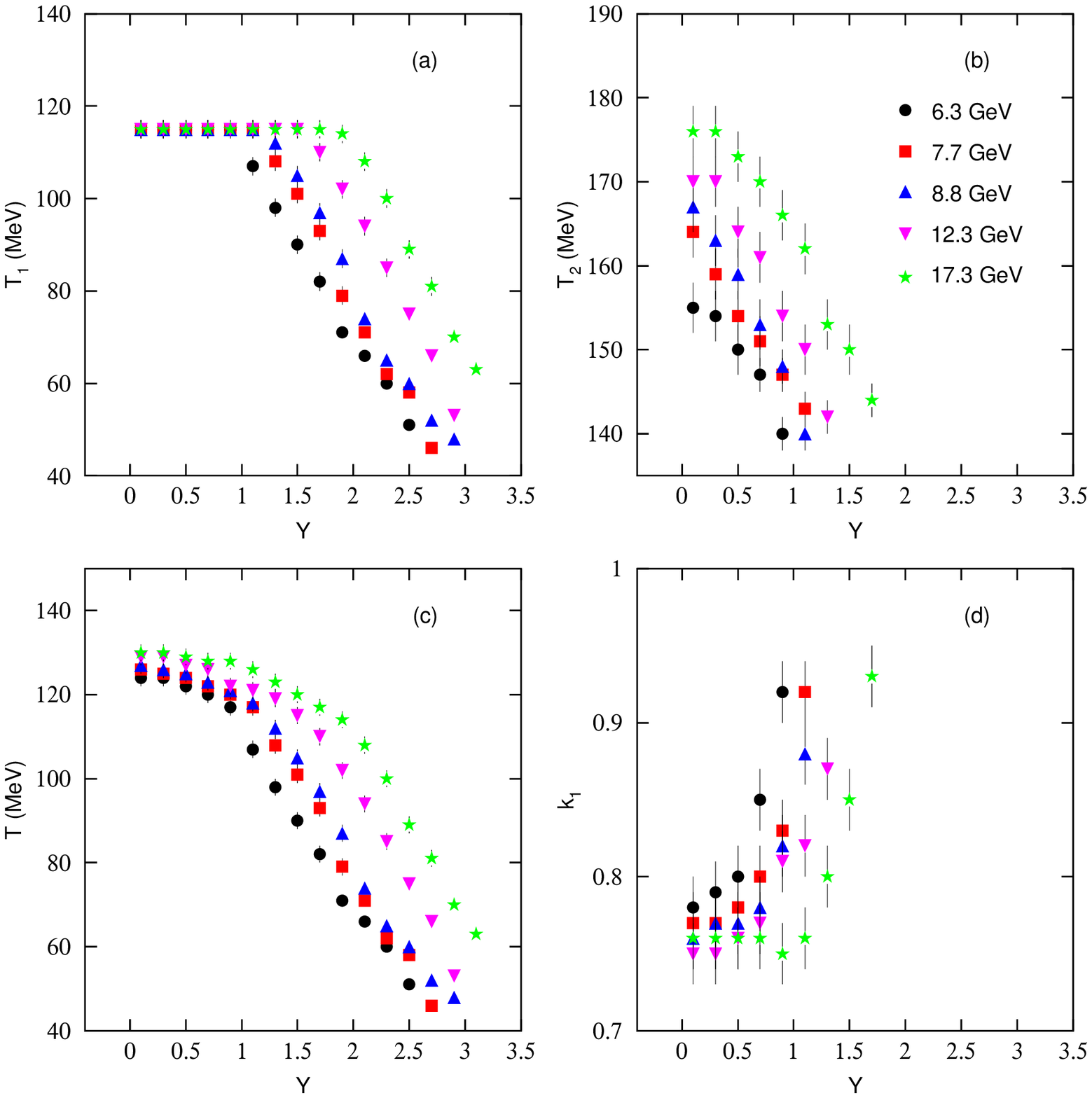}
\end{center}
\vskip-.0cm {\small Figs. 7. (color online) Correlations between
(a) $T_1$ and $Y$, (b) $T_2$ and $Y$, (c) $T$ and $Y$, as well as
(d) $k_1$ and $Y$ in pp collisions at different $\sqrt{s_{\rm
NN}}$. The results corresponding to different $\sqrt{s_{\rm NN}}$
are represented by different symbols marked in the panels.}
\end{figure*}

\begin{figure*}[htb]
\vskip-.0cm
\begin{center}
\includegraphics[width=13.cm]{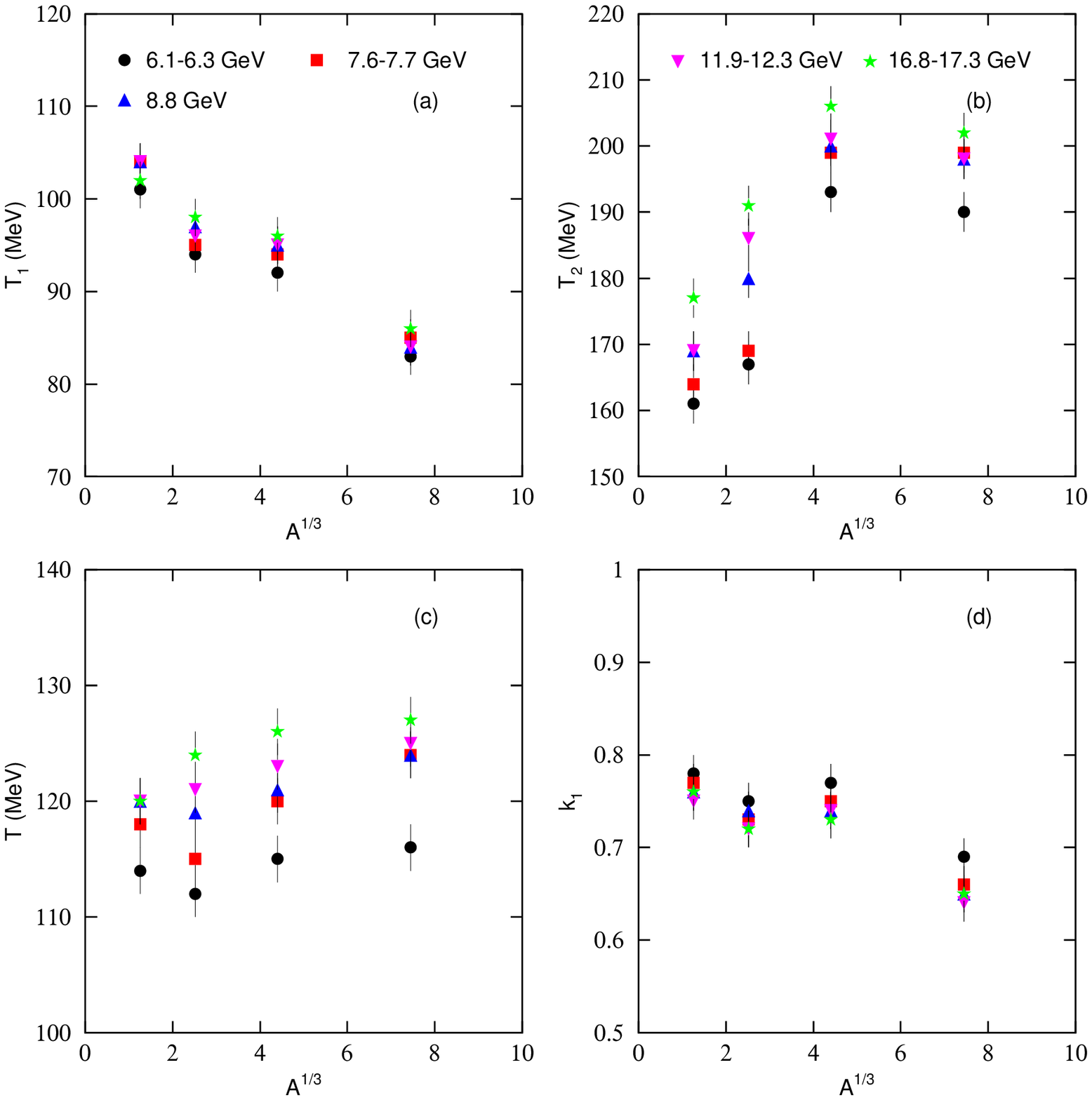}
\end{center}
\vskip.0cm {\small Figs. 8. (color online) Correlations between
(a) $T_1$ and $A^{1/3}$, (b) $T_2$ and $A^{1/3}$, (c) $T$ and
$A^{1/3}$, as well as (d) $k_1$ and $A^{1/3}$ at different
$\sqrt{s_{\rm NN}}$.}
\end{figure*}

\begin{figure*}[htb]
\begin{center}
\includegraphics[width=13.cm]{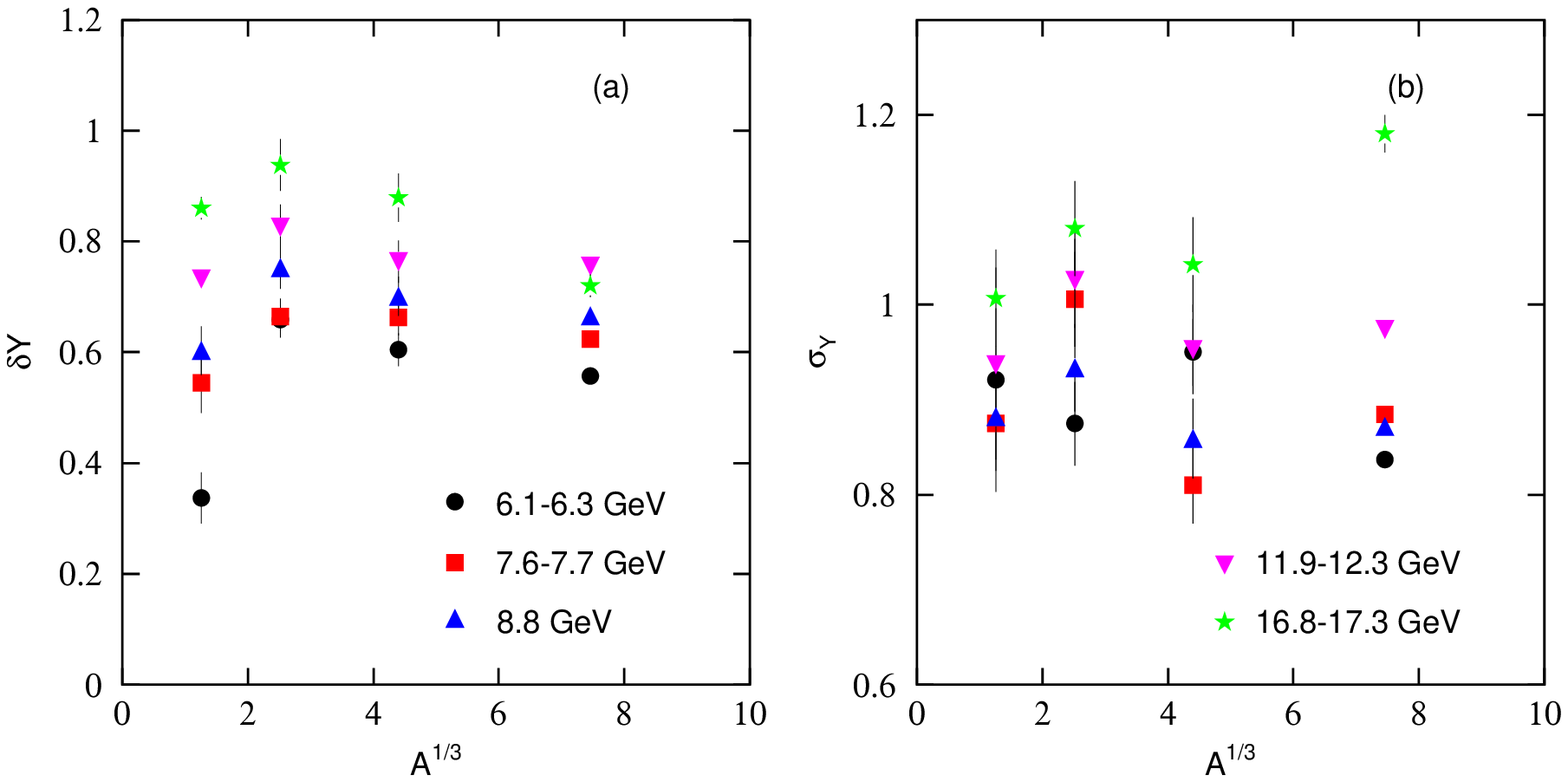}
\end{center}
\vskip.0cm {\small Figs. 9. (color online) Correlations between
(a) $\delta Y$ and $A^{1/3}$, as well as (b) $\sigma_Y$ and
$A^{1/3}$ at different $\sqrt{s_{\rm NN}}$.}
\end{figure*}

Now we move to a discussion of the properties of parameters
obtained from Figs. 1--3 and listed in Tables 2--4. To show
clearly the pictures for the dependences of parameters on $Y$, $A$
and $\sqrt{s_{\rm NN}}$, Fig. 7 shows the correlations between (a)
$T_1$ and $Y$, (b) $T_2$ and $Y$, (c) $T$ and $Y$, as well as (d)
$k_1$ and $Y$ in pp collisions at different $\sqrt{s_{\rm NN}}$.
Fig. 8 shows the correlations between (a) $T_1$ and $A^{1/3}$, (b)
$T_2$ and $A^{1/3}$, (c) $T$ and $A^{1/3}$, as well as (d) $k_1$
and $A^{1/3}$ at different $\sqrt{s_{\rm NN}}$. Fig. 9 shows the
correlations between (a) $\delta Y$ and $A^{1/3}$, as well as (b)
$\sigma_Y$ and $A^{1/3}$ at different $\sqrt{s_{\rm NN}}$. One can
see from Fig. 7 that, in pp collisions, $T_1$ does not change in
central rapidity region and decreases in forward rapidity region,
$T_2$ and $T$ decrease, and $k_1$ increases with increase of $Y$.
From Fig. 8 one can see that, from pp to central Pb-Pb collisions,
$T_1$ decreases, $T_2$ increases and then saturates, $T$ increases
slightly, and $k_1$ decreases. From Fig. 9 one can see that
$\delta Y$ and $\sigma_Y$ do not change obviously. Meanwhile, from
Figs. 7--9 one can see that, with increase of $\sqrt{s_{\rm NN}}$,
$T_1$ does not change in central rapidity region and increases in
forward rapidity region (or $T_1$ increases slightly), $T_2$ and
$T$ increase, $k_1$ decreases, and $\delta Y$ and $\sigma_Y$
increase.

The observed dependencies of the parameters can be treated as
follows in the sense of possible scenarios of the particle
production process.

The observations in Fig. 7, namely the increase of the temperature
parameters $T_1$, $T_2$, and then the increase of the effective
temperature $T$ being their combination, with decrease of the
rapidity $Y$ can be explained by an increase of the excitation
energy due to the increasing energy deposition by the system as
one moves from the large absolute rapidities (above unity, the
forward rapidity region) to the lower ones (central region). In
the forward rapidity region it is natural to assume the particles
having a shorter mean-free path as less energy is deposited, while
in the central rapidity region, the particles can be considered to
scatter with larger mean-free path as getting larger excitation
energy. This picture can be associated with the liquid-like state
in the former case against the gas-like state in the latter one.
Within this consideration, the first component characterized by
$T_1$ temperature can be assigned to the liquid-like state, and
the second component with $T_2$ temperature to the gas-like state.
Then, the temperature of a phase transition from the liquid-like
state to the gas-like one gets 115 MeV (see Fig. 7(a)) with no
flow effect being excluded. This also makes it understandable the
$T_1$ parameter to be of a constant value as one reaching a
critical temperature value. One however should note that in the
central rapidity region, the mixture of the gas-like and
liquid-like states is expected, as it is seen from the
$T$-parameter behavior.

The system-size dependences shown in Fig. 8 find their following
explanations. For the liquid-like state, one expects lower
temperature in larger systems due to the energy loss, as it is
observed in Fig. 8(a). On the contrary, for the gas-like
component, more energy deposition should be expected in large
system due to nuclear stopping and hence higher temperature as it
is seen in Fig. 8(b).

The behavior of the effective temperature parameter $T$ is a
combination of the contributions of $T_1$, $T_2$ and the rate
$k_1$. From Figs. 7 and 8, one can see that $T$ increases and
$k_1$ decreases as one moves from the forward region to the
central one and as the collision energy and the system size
increase. These dependences can be explained by the general
increase of the energy deposition. Let us remind that the $T$ is
the effective temperature which means it does not exclude the
particle-flow effect. As soon as one excludes the flow effect, the
lower temperature value can be obtained characterizing the
interacting system a freeze-out, the so-called freeze-out
temperature.

The dependences observed in Fig. 9 can be treated as the following
features of the particle production process. Assuming the binary
nucleon-nucleon collisions being the main internal process not
only in small system as pp but also in such a large system as
Pb-Pb is, the rapidity shift independence of the system size is
expected as it is generally confirmed by the observation in Fig.
9. In the meantime, the increase of the rapidity shift with the
collision energy would be naturally explained by the increase of
the penetrating power.

It should be noted that the functions used in the above
discussions are only one choice to fit the experimental $p_{\rm
T}$ or $m_{\rm T}$ and $Y$ spectra. One can also choose other
functions to fit the same experimental spectra and to obtain
properties of other parameters. One can consider the Tsallis
distribution to fit the transverse spectra as widely used, see
e.g. ref. [8]. However, the effective temperature considered in
the present work and one used in the Tsallis approach cannot be
compared directly as being different parameters.
\\

{\section{Conclusions}}

We summarize here our main observations and conclusions.

(a) We have used the hybrid model to fit the $p_{\rm T}$ or
$m_{\rm T}$ and $Y$ spectra of $\pi^-$ produced in pp and central
Be-Be, Ar-Sc, as well as Pb-Pb collisions at the SPS at its BES
energies which cover an energy range from 6.1 to 17.3 GeV. For the
$p_{\rm T}$ or $m_{\rm T}$ spectra, the (two-component) standard
distribution is used. For the $Y$ spectra, the double-Gaussian
distribution is used. The model results are in good agreement with
the experimental data of the NA61/SHINE and NA49 Collaborations.
All parameter values and event patterns extracted from the fits
reflect the properties of interaction system at the stage of
kinetic freeze-out.

(b) The extracted parameters show abundant features. In pp
collisions, the distribution first component effective temperature
$T_1$ does not change in central rapidity region and decreases in
forward rapidity region, the distribution second component
effective temperature $T_2$ and the combined effective temperature
$T$ decrease, and the fraction parameter $k_1$ increases with
increase of $Y$. From pp to central Pb-Pb collisions, $T_1$
decreases, $T_2$ increases and then saturates, $T$ increases
slightly, $k_1$ decreases, and $\delta Y$ and $\sigma_Y$ do not
change obviously. Meanwhile, with increase of $\sqrt{s_{\rm NN}}$,
$T_1$ does not change in central rapidity region and increases in
forward rapidity region (or $T_1$ increases slightly), $T_2$ and
$T$ increase, $k_1$ decreases, and $\delta Y$ and $\sigma_Y$
increase.

(c) The two types of event patterns are extracted in pp
collisions. The fine patterns are extracted from the $p_{\rm T}$
spectra which depend on rapidity, and the other, non-fine patterns
are from the rapidity-independent $m_{\rm T}$ spectra which is
only measured at mid-rapidity. Both the event patterns in
$\beta_x$-$\beta_y$-$\beta_z$ space are spherical, though
$\sqrt{\overline{\beta_{x,y}^2}}$ ($\sqrt{\overline{\beta_z^2}}$)
in fine event pattern is equal to or larger (less) than that in
non-fine event pattern. Both the event patterns in
$p_x$-$p_y$-$p_z$ space are rough cylindrical, though the size of
fine event pattern is less than that of non-fine event patterns.
Both the event patterns in $Y_1$-$Y_2$-$Y$ space are rough
rhombohedral, though the size of fine event pattern in transverse
plane is larger than that of non-fine event patterns.

(d) From pp to central Pb-Pb collisions, there is no obvious
change in the shape and size of non-fine event patterns. With
increase of $\sqrt{s_{\rm NN}}$, in both the three-dimensional
velocity and rapidity spaces, the size of event pattern in
transverse plane decreases and that in longitudinal direction
increases. Meanwhile, in the three-dimensional momentum space, the
size of event pattern in transverse plane does not change
obviously, and that in longitudinal direction increases obviously.
These observations are useful in the understanding of particle
production in different types of collisions such as the collisions
at the BES at the Relativistic Heavy Ion Collider.
\\

{\bf Acknowledgments}

This work was supported by the National Natural Science Foundation
of China under Grant Nos. 11575103 and 11747319, the Shanxi
Provincial Natural Science Foundation under Grant No.
201701D121005, and the Fund for Shanxi ``1331 Project" Key
Subjects Construction.
\\

\end{multicols}
\end{document}